\documentclass[a4paper,10pt]{article}

\pdfoutput=1  

\usepackage{amsmath,amssymb,mathtools}     
\usepackage{color}
\usepackage{graphicx}
\usepackage{subfigure}
\usepackage{cite}                
\usepackage{hyperref}            
\usepackage{multirow,makecell}   
\usepackage{textcomp}
\usepackage{wasysym}
\usepackage[text={17.1cm,24.5cm},centering]{geometry}

\numberwithin{equation}{section}   

\def \be {\begin{equation}}
\def \ee {\end{equation}}
\def \ba {\begin{array}}
\def \ea {\end{array}}
\def \bea{\begin{eqnarray}}
\def \eea{\end{eqnarray}}
\def \nn {\nonumber}

\newcommand{\eq}[1]{(\ref{#1})}

\def \a {\alpha}
\def \b {\beta}

\def \D {\Delta}
\def \e {\epsilon}

\def \s {\sigma}

\def \r {\rho}

\def \vph {\varphi}

\def \mA {\mathcal A}
\def \mB {\mathcal B}
\def \mC {\mathcal C}
\def \mD {\mathcal D}
\def \mE {\mathcal E}
\def \mF {\mathcal F}

\def \mH {\mathcal H}
\def \mI {\mathcal I}
\def \mJ {\mathcal J}
\def \mK {\mathcal K}
\def \mL {\mathcal L}
\def \mM {\mathcal M}
\def \mN {\mathcal N}
\def \mO {\mathcal O}
\def \mP {\mathcal P}
\def \mQ {\mathcal Q}
\def \mR {\mathcal R}

\def \mX {\mathcal X}

\def \p {\partial}
\def \f {\frac}

\def \mc {\mathcal}

\def \lt {\left}
\def \rt {\right}

\def \sr {\sqrt}
\def \td {\tilde}

\def \inf {\infty}

\def \lag {\langle}
\def \rag {\rangle}

\def \ep {\mathrm{e}}
\def \ii {\mathrm{i}}

\def \tr {\textrm{tr}}

\def \and {{\textrm{and}}}

\def \CFT {{\textrm{CFT}}}

\def \Z {{\textrm{Z}}}

\def \cyl {{\textrm{cyl}}}

\begin{document}

\title{\textbf{Subsystem eigenstate thermalization hypothesis for entanglement entropy in CFT}}
\author{
Song He$^{1,2}$\footnote{hesong17@gmail.com}~,
Feng-Li Lin$^{3}$\footnote{fengli.lin@gmail.com}~
and
Jia-ju Zhang$^{4,5}$\footnote{jiaju.zhang@mib.infn.it}
}
\date{}

\maketitle
\vspace{-10mm}
\begin{center}
{\it
$^{1}$Max Planck Institute for Gravitational Physics (Albert Einstein Institute),\\
Am M\"uhlenberg 1, 14476 Golm, Germany\\\vspace{1mm}
$^{2}$CAS Key Laboratory of Theoretical Physics, Institute of Theoretical Physics, Chinese Academy of Sciences,\\
55 Zhong Guan Cun East Road, Beijing 100190, China\\\vspace{1mm}
$^{3}$Department of Physics, National Taiwan Normal University, \\
No. 88, Sec. 4, Ting-Chou Rd., Taipei 11677, Taiwan\\\vspace{1mm}
$^{4}$Dipartimento di Fisica, Universit\'a degli Studi di Milano-Bicocca, Piazza della Scienza 3, I-20126 Milano, Italy\\\vspace{1mm}
$^{5}$INFN, Sezione di Milano-Bicocca, Piazza della Scienza 3, I-20126 Milano, Italy\\\vspace{1mm}
}
\vspace{10mm}
\end{center}

\begin{abstract}

  We investigate a weak version of subsystem eigenstate thermalization hypothesis (ETH) for a two-dimensional large central charge conformal field theory by comparing the local equivalence of high energy state and thermal state of canonical ensemble.  We evaluate the single-interval R\'enyi entropy and entanglement entropy for a heavy primary state in short interval expansion. We verify the results of R\'enyi entropy by two different replica methods. We find nontrivial results at the eighth order of short interval expansion, which include an infinite number of higher order terms in the large central charge expansion.  We then evaluate the relative entropy of the reduced density matrices to measure the difference between the heavy primary state and thermal state of canonical ensemble, and find that the aforementioned nontrivial eighth order results make the relative entropy unsuppressed in the large central charge limit. By using Pinsker's and Fannes-Audenaert inequalities, we can exploit the results of relative entropy to yield the lower and upper bounds on trace distance of the excited-state and thermal-state reduced density matrices. Our results are consistent with subsystem weak ETH, which requires the above trace distance is of power-law suppression by the large central charge. However, we are unable to pin down the exponent of power-law suppression.  As a byproduct we also calculate the relative entropy to measure the difference between the reduced density matrices of two different heavy primary states.

\end{abstract}

\baselineskip 18pt
\thispagestyle{empty}
\newpage

\tableofcontents


\section{Introduction}

According to eigenstate thermalization hypothesis (ETH) \cite{Deutsch:1991,Srednicki:1994}, a highly excited state of a chaotic system behaves like a high engergy microcanonical ensemble thermal state. More precisely, it states that (i) the diagonal matrix element $\mA_{\alpha\alpha}$ of a few-body operator $\mA$ with respect to the energy eigenstate $\alpha$ change slowly with the state in a way of suppression by the exponential of the system size; (ii) the off-diagonal element $\mA_{\alpha\beta}$ is much smaller than the diagonal element by the factor of exponential of the system size.  This will then yield that the expectation value of the few-body observable in a generic state $|\phi\rag$ behaves like the ones in the microcanonical ensemble
\be \label{localETH}
\lag \mA \rag_\phi - \lag \mA \rag_{{E}} \sim \ep^{-\mO(S(E))},
\ee
where the subscript $E$ denotes microcanonical ensemble state with energy $E$ and $S(E)$ is the system entropy.

Recently, a new scheme called subsystem ETH has been proposed in \cite{Lashkari:2016vgj,Dymarsky:2016aqv},  in contrast to the old one which is then called local ETH. The subsystem ETH states that the reduced density matrix $\r_{A,\phi}$ of a subregion $A$ for a high energy primary eigenstate $|\phi\rag$ are universal to the reduced density matrix $\r_{A,E}$ for microcanonical ensemble thermal state with some engery $E$ up to exponential suppression by order of the system entropy, i.e.,
\be\label{subETH}
t(\rho_{A,\phi},\rho_{A,E}) \sim \ep^{-\mathcal{O}(S(E))},
\ee
where $t(\rho_{A,\phi},\rho_{A,E})$ denotes the trace distance between $\rho_{A,\phi}$ and $\rho_{A,E}$.  As the subsystem ETH is a statement regarding the reduced density matrices, their derived quantities such as correlation functions, entanglement entropy and R\'enyi entropy should also satisfy some sort of the subsystem ETH. In this sense, the subsystem ETH is strongest form of ETH, i.e., stronger than local ETH.

  The above discussions of ETH are all based on the comparison of energy eigenstate and the microcanonical ensemble state. In \cite{Iyoda:2016}, there are discussions of generalizing the ETH to the comparison with the canonical ensemble state, based on the observation of the local equivalence between the canonical and microcanonical states \cite{Tasaki:2016}. In this case, the energy eigenstate and canonical ensemble thermal state should also locally alike in the thermodynamic limit. This was called weak ETH in \cite{Iyoda:2016}, in which, however, there is no exact bound on weak ETH for general cases. Despite that, some of the results in \cite{Tasaki:2016} showed that the thermal states of canonical and microcanonical  ensembles are locally equivalent up to power-law suppression of the dimension of Hilbert space. Based on all the above, we may expect that the weak ETH will at least yield
\be\label{weakETH}
\lag \mA \rag_\phi - \lag \mA \rag_\b \sim [S(\b)]^{-\tilde{a}}, ~~ t(\rho_{A,\phi},\rho_{A,\b}) \sim [S(\b)]^{-\tilde{b}},
\ee
where the subscript $\b$ denotes canonical ensemble with inverse temperature $\b$, $S(\b)$ is the canonical ensemble entropy, and $\tilde{a}$ and $\tilde{b}$ are some positive real numbers of order one. More specifically, it was argued in \cite{Srednicki-1} that $\tilde{a}=1$ and was verified in \cite{Alba} by numerical simulation for some integrable model. However, for generic models it was mathematically shown that $\tilde{a}=1/4$ \cite{Iyoda:2016}. We expect $\tilde{b}$ should also behave similarly. We will call the 2nd inequality in \eq{weakETH} the subsystem weak ETH, which inherits both the subsystem ETH and weak ETH.

For a conformal field theory (CFT) with there are infinite degrees of freedom, which is in some sense the thermodynamic limit, as required for the local equivalence between canonical and microcanonical states.  Moreover, the nonlocal quantities like entanglement entropy and R\'enyi entropy do not necessarily be exponentially suppressed \cite{Lashkari:2016vgj,Dymarsky:2016aqv}. In fact the primary excited state R\'enyi entropy in two-dimensional (2D) CFT is not exponentially suppressed in the large central charge limit \cite{Lashkari:2016vgj,Lin:2016dxa}.  All of these motivate us to check the validity of weak ETH for a 2D CFT of large central charge $c$.

In this paper we investigate the validity of the subsystem weak ETH \eq{weakETH} for a 2D  large $c$ CFT. In this case, the worldsheet description of an excited state $|\phi\rag$ for a CFT living on a circle of size $L$ corresponds to an infinitely long cylinder of spatial period $L$ capped by an operator $\phi$ at each end. On the other hand, the thermal state of a CFT living on a circle with temperature $T$ has its worldsheet description as a  torus with temporal circle of size $\beta=1/T$. In the high temperature limit with $L \gg \beta$, the torus is approximated by a horizontal cylinder. Naively the vertical and horizontal cylinders should be related by Wick rotation and can be compared after taking care of the capped states. This is indeed what have been done in \cite{Fitzpatrick:2014vua,Fitzpatrick:2015zha} by comparing the two-point functions two light operators of large $c$ CFT, and in \cite{Asplund:2014coa,Caputa:2014eta} for the single-interval entanglement entropy. These comparisons all show that the subsystem weak ETH holds. However, in \cite{Lashkari:2016vgj,Lin:2016dxa} the one-interval R\'enyi entropy for small interval of size $\ell\ll L$ are compared by $\ell$ expansion up to order  $\ell^6$ and it was found that one cannot find a universal relation between $\beta$ and $L$ to match the excited-state R\'enyi entropy with the thermal one in the series expansion of $\ell$.\footnote{Note that in \cite{Lashkari:2016vgj} no large $c$ is required, and they just require the excited state to be heavy.}

Moreover, in the context of AdS/CFT correspondence \cite{Maldacena:1997re,Gubser:1998bc,Witten:1998qj,Aharony:1999ti}, a large $N$ CFT is dual to the AdS gravity of large AdS radius, and so the subsystem ETH implies that the backreacted geometry by the massive bulk field is approximately equivalent to the black hole geometry for the subregion observer. Especially, for AdS$_3$/CFT$_2$ the CFT has infinite dimensional conformal symmetries as the asymptotic symmetries of AdS space \cite{Brown:1986nw}, along with ETH it could imply that the infinite varieties of Ba\~nados geometries \cite{Banados:1998gg} dual to the excited CFT states are universally close to the BTZ black hole \cite{Banados:1992wn}. Although the Newton constant $G_N$ may get renormalized in the $1/c$ perturbation theory, and then obscure the implication of our results to the above issue, we still hope that our results can provide as the stepping stone for the  further progress.

As discussed in \cite{Lashkari:2016vgj}, the validity of subsystem ETH depends on how the operator product expansion (OPE) coefficients scale with the conformal dimension of the eigen-energy operator in the thermodynamic limit.
This means that the subsystem ETH could be violated for some circumstances.
In this paper we continue to investigate the validity of ETH for a 2D large $c$ CFT by more extensive calculations, and indeed find the surprising results.
According to Cardy's formula \cite{Cardy:1986ie} the thermal entropy is proportional to the central charge $c$, and so we just focus on how various quantities behave in large $c$ limit.
We calculate the entanglement entropy and R\'enyi entropy up to order $\ell^8$ in the small $\ell$ expansion.
We then find that there appear subleading corrections of $1/c$ expansion at the order $\ell^8$.
Because the appearance of these subleading corrections at order $\ell^8$ is quite unexpected, we solidify the results by adopt two different method to calculate them. These methods are (i) the OPE of  twist operators  \cite{Calabrese:2004eu,Headrick:2010zt,Calabrese:2010he,Chen:2013kpa} on cylinder, or equivalently on complex plane; and (ii) $2n$-point correlation function on complex plane \cite{Alcaraz:2011tn,Berganza:2011mh,Lashkari:2014yva,Lashkari:2015dia,Sarosi:2016oks,Sarosi:2016atx,Ruggiero:2016khg}. By both the two methods we get the same results.
Moreover, we turn the comparison of the entanglement entropy into the relative entropy between reduced density matrices for excited state and thermal state by the modular Hamiltonian argument in \cite{Lashkari:2016vgj}.
Then, the above discrepancy yields that the relative entropy is of order $c^0$.

   Based on the above results, we use the Fannes-Audenaert inequality \cite{Fannes1973,Audenaert:2006} and Pinsker's inequality by relating the trace distance to entanglement entropy or relative entropy, to argue how the trace distance of the reduced density matrices of the excited and thermal states scales with the large $c$. Our results are consistent with the subsystem weak ETH. However, we lack of further evidence to pin-down the exact power-law suppression, i.e., unable to obtain the exponent $\tilde{b}$ in \eq{weakETH}.

    Finally, using the replica method based on evaluating the multi-point function on a complex plane, as a byproduct of this project we explicitly calculate the relative entropy to measure the difference between the reduced density matrices of two different heavy primary states.

The rest of the paper is organized as follows.
In section~\ref{sec2} we briefly review the known useful results about R\'enyi entropy, entanglement entropy and relative entropy, and also evaluate the relative entropy between chiral vertex state and thermal state in 2D free massless scalar theory.
In section~\ref{sec3} we calculate the excited state R\'enyi entropy in two different replica methods in the short interval expansion.
Using these results, in section~\ref{sec4} we check the subsystem weak ETH. We find that the validity of the subsystem weak ETH depend only on whether the large but finite effective dimension of the reduced density matrix exists and if yes how it scales with $c$.
In section~\ref{sec5} we evaluate the relative entropy of the reduced density matrices to measure the difference of two heavy primary states.
Finally, we conclude our paper in section~\ref{sec6}.
Moreover, in appendix~\ref{vcf} we give some details of the vacuum conformal family;
in appendix~\ref{opests} we give the results of OPE of twist operators, including both the review of the formulism and some new calculations; and in appendix~\ref{usf} we list some useful summation formulas.

\section{Relative entropy and ETH}\label{sec2}

In this section we will briefly review the basics of relative entropy and then ETH. In the next section, the relative entropy between the reduced density matrices of heavy state and thermal state will be evaluated for large central charge 2D CFT to check the validity of ETH. We will end this section by calculating the relative entropy of a toy example CFT, namely the 2D massless scalar.

\subsection{Relative entropy}

Given a quantum state of a system denoted by the density matrix $\rho$, then the reduced density matrix on some region $A$ is given by
\be
\r_A=\tr_{A^c}\r.
\ee
where $A^c$ is the complement of $A$.  One can then define the R\'enyi entropy
\be
S_{A,n}=-\f{1}{n-1}\log \tr_A\r_A^n,
\ee
and the entanglement entropy
\be
S_A=-\tr_A(\r_A\log\r_A),
\ee
which is formally equivalent to taking $n\to1$ limit of the R\'enyi entropy.

In this work we will focus on the holomorphic sector of a 2D CFT of central charge $c$, and the R\'enyi entropy for a region $A$ of size $\ell$, i.e., $A=[-\ell/2,\ell/2]$, for a vacuum state it is known \cite{Calabrese:2004eu}
\be \label{SnL}
S_{n,L}=\f{c(n+1)}{12n}\log \Big( \f{L}{\pi \e} \sin \f{\pi\ell}{L} \Big),
\ee
where $L$ is the size of the spatial circle on which the CFT lives. Similarly, the R\'enyi entropy of a thermal state with temperature $1/\beta$ for a CFT living on a infinite straight line is
\be \label{Snb}
S_{n,\b}=\f{c(n+1)}{12n}\log \Big( \f{\b}{\pi \e} \sinh \f{\pi\ell}{\b} \Big).
\ee
Taking $n \to 1$ limit, one can get the corresponding entanglement entropy straightforwardly. For simplicity, in this paper we only consider the contributions from the holomorphic sector of CFT, and the anti-holomorphic sector can be just added for completeness without complication. Also we will not consider the subtlety due to the boundary conditions imposed on the entangling surface \cite{Ohmori:2014eia,Cardy:2016fqc}.

For subsystem ETH it is to compare the reduced density matrices between a heavy state and the excited state, and in this paper we consider the R\'enyi entropy difference, the entanglement entropy difference, and the relative entropy, as well as the trace distance.
The relative entropy is defined as
\be
S(\r'_A\|\r_A) = \tr_A \r'_A\log\r'_A - \tr_A \r'_A\log\r_A.
\ee
where $\r_A$ and $\r'_A$ are the reduced density matrixes over region $A$ for state $\r$ and $\r'$, respectively.  Note that the relative entropy is not symmetric for its two arguments, i.e., $S(\r'_A\|\r_A) \ne S(\r_A\|\r'_A)$. One may define the symmetrized relative entropy
\be
S(\r'_A,\r_A) =  S(\r'_A\|\r_A) + S(\r_A\|\r'_A),
\ee
to characterize the difference of the two reduced density matrices, but one should be aware that it is not a ``distance''.\footnote{We thank the anonymous referee for pointing this out to us.}

One can also express the relative entropy as follows:
\be \label{e39}
S(\r'_A\|\r_A) = \lag H_A \rag_{\r'} - \lag H_A \rag_{\r} -S'_A +S_A.
\ee
where the modular Hamiltonian $H_A$ is defined by
\be
H_A \equiv -\log \rho_A.
\ee
The modular Hamiltonian is in general quite nonlocal and known only for some special cases \cite{Casini:2011kv,Wong:2013gua,Lashkari:2015dia,Cardy:2016fqc}.
One of these cases fitted for our study in this paper is just the case considered for the R\'enyi entropy \eq{Snb} of a thermal state, and the modular Hamiltonian is given by \cite{Wong:2013gua}
\be \label{e40}
H_{A,\b} = - \f{\b}{\pi} \int_{-\ell/2}^{\ell/2} dx \; \f{\sinh\f{\pi(\ell-2x)}{2\b}\sinh\f{\pi(\ell+2x)}{2\b}}{\sinh\f{\pi\ell}{\b}} \; T(x)
\ee
where $T(x)$ is the holomorphic sector stress tensor of the 2D CFT.

In this paper we will check the subsystem weak ETH for a normalized highly and globally excited state\footnote{In this paper, we mainly focus on global excited states which are quite different from so called locally excited states studied in, for examples, \cite{Nozaki:2014hna,He:2014mwa,Guo:2015uwa,Chen:2015usa}.} created by a (holomorphic) primary operator $\phi$ of conformal weight  $h_\phi=c\e_\phi$ acting on the vacuum, i.e., $|\phi\rag =\phi(0) |0\rag$. 
The first step to proceed the comparison for checking ETH is to make sure the excited state and the thermal state have the same energy, and this then requires
\be \label{e41}
\lag\phi|T|\phi\rag_L = \lag T \rag_\b.
\ee
The right hand side of the above equation is just the Casimir energy of the horizontal worldsheet cylinder
\be
\lag T \rag_\b = -\frac{\pi^2 c}{6 \beta^2}.
\ee
and the left hand side is given by (\ref{phikphi}). Thus, \eq{e41} yields a relation between the inverse temperature $\b$ and the conformal weight  $h_\phi$ (or $\e_\phi)$ \cite{Fitzpatrick:2014vua,Fitzpatrick:2015zha}
\be \label{e47}
\b=\f{L}{\sr{24\e_\phi-1}}.
\ee
Moreover, the relation \eq{e41} and (\ref{e40}) ensure $\lag H_{A,\b} \rag_{\phi} = \lag H_{A,\b} \rag_{\b}$ so that \cite{Lashkari:2016vgj}
\be \label{LDL}
S(\r_{A,\phi}\|\r_{A,\b}) = -S_{A,\phi} + S_{A,\b}.
\ee

\subsection{ETH}\label{ETH-I}

   ETH states that a highly excited state of a chaotic system behaves thermally. One way to formulate this is to compare the expectation values of few-body operators for high energy eigenstate and the thermal state, as explicitly formulated in \eq{localETH}. This is called the local ETH in \cite{Dymarsky:2016aqv} in contrast to a stronger statement called subsystem ETH proposed therein, which is formulated as in \eq{subETH} by comparing the reduced density matrices, i.e., requiring that trace distance between the two reduced states should be exponentially suppressed by the system entropy.  The trace distance for two reduced density matrices $\r_A'$, $\r_A$ is defined as
\be\label{traceD}
t(\r_A',\r_A) = \f12 \tr_A | \r_A'-\r_A |,
\ee
and by definition $0 \leq t(\r_A',\r_A) \leq 1$.

   In the paper we compare the energy eigenstate and canonical ensemble state, and so it is about the local weak ETH and subsystem weak ETH. As the subsystem weak ETH is stronger than local weak ETH, it could be violated for the system of infinite number of degrees of freedom.   However, we do not directly calculate the trace distance but the R\'enyi entropies and entanglement entropies for both heavy state and thermal state of canonical ensemble. After doing this, we can then use some inequalities to constrain the trace distance with the difference of the R\'enyi entropies or relative entropy, thus check the validity of subsystem weak ETH.

   Here are three such kinds of inequality.  First, the Fannes-Audenaert inequality \cite{Fannes1973,Audenaert:2006} relating the difference of entanglement entropy, $\D S_A := S_{A,\phi} -S_{A,\b}$ to the trace distance $t:=t(\r_{A,\phi},\r_{A,\b})$ as follows:
\be \label{FAI}
|\D S_A| \leq t \log (d-1) +h,
\ee
with $h=-t\log t -(1-t)\log(1-t)$ and $d$ being the dimension of  Hilbert space $\mH_A$ for the effective degrees of freedom in subsystem $A$. On the other hand, there is the Audenaert inequality for the R\'enyi entropy of order $0<n<1$ \cite{Audenaert:2006}
\be\label{AI}
|\D S_n| \leq \f{1}{1-n} \log[ (1-t)^n + (d-1)^{1-n}t^n ],
\ee
with $\D S_n:=S_{n,\phi} -S_{n,\b}$.
Both the right hand sides of \eq{FAI} and \eq{AI} are vanishing at $t=0$, are $\log(d-1)$ at $t=1$, monotonically increase at $0<t<1-\f1{d}$, monotonically decrease at $1-\f1{d}<t<1$, and have a maximal value $\log d$ at $t=1-\f1{d}$. Since $d$ is very large, the right hand sides of \eq{FAI} and \eq{AI} are approximately monotonically increase at $0<t<1$.
Finally, we also need Pinsker's inequality to give upper bound on trace distance by the square root of relative entropy, i.e.
\be \label{PI}
t \leq \sr{\f12 S(\r_{A,\phi}\|\r_{A,\b})}.
\ee

By using \eq{FAI} we see that $|\D S_A|$ gives the tight lower-bound on the trace distance if the $d$ is finite, thus the validity of subsystem weak ETH can be pin down by the scaling behavior of $|\D S_A|$ with respect to the system entropy.
This is no longer true if $d$ is infinite as one would expect for generic quantum field theories, then both \eq{FAI} and \eq{AI} are trivially satisfied and can tell no information about the trace distance \cite{Lashkari:2016vgj,Dymarsky:2016aqv}. However, it is a subtle issue to find out how the effective dimension $d$ of the reduced density matrix scale with the large $c$ and if it is finite once a UV cutoff is introduced.  We will discuss in more details in section~\ref{sec4}.

\subsection{A toy example}

We now apply the above formulas to a toy 2D CFT, the massless free scalar. This CFT has central charge $c=1$ so that it makes less sense to check subsystem weak ETH. Despite that, we will still calculate the relative entropy between excited state and thermal state, and the result can be compared to the large $c$ ones obtained later.

\begin{figure}[htbp]
  \centering
  \includegraphics[width=0.618\textwidth]{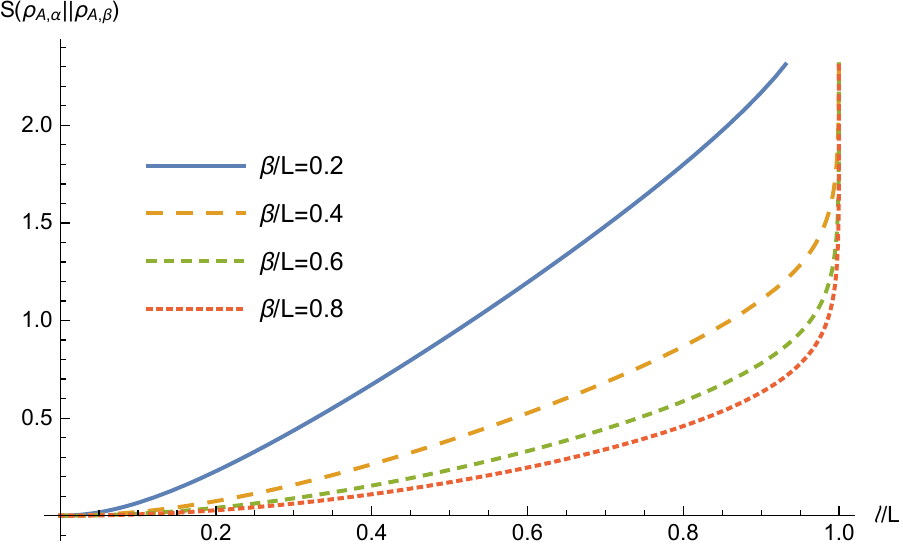}\\
  \caption{Relative entropy \eq{j45} as a function of $\ell/L$ in comparing the reduced density matrices for chiral primary state and thermal state of 2D massless scalar. Note that the higher the temperature is, the larger the relative entropy becomes.}\label{ms}
\end{figure}

Let the massless scalar denoted by $\vph$, and from it we can construct the chiral vertex operator \cite{DiFrancesco:1997nk}
\be
V_\a(z) = \ep^{\ii\a\vph(z)},
\ee
with conformal weight
\be
h_\a=\f{\a^2}{2}.
\ee
Choosing $\a\gg 1$ we can create the highly excited state as follows:
\be
|V_\a\rag = V_\a(0)|0\rag.
\ee
The R\'enyi entropy for the state $|V_\a\rag$ was calculated before in \cite{Berganza:2011mh,Lashkari:2014yva,Ruggiero:2016khg}, and the result is the same as \eq{SnL} for the vacuum state, no matter what the value $\a$ is.  Thus, the relative entropy \eq{LDL} can be obtained straightforwardly and the result is
\be \label{j45}
S(\r_{A,\a}\|\r_{A,\b}) = \f{1}{6} \log \f{\b\sinh\f{\pi\ell}{\b}}{L\sin\f{\pi\ell}{L}}.
\ee
In Fig.~\ref{ms} we have plotted the results as a function of $\ell/L$ for various $\b/L$.  We see that the relative entropy is overall larger for heavier excited state. Note that $\a$ appears in \eq{j45} implicitly through
\be
\b = \f{L}{\sr{12\a^2-1}}.
\ee

\section{Excited state R\'enyi entropy}\label{sec3}

We now consider the 2D CFT with large central charge, which can be also thought as dual CFT of AdS$_3$. We aim to calculate the R\'enyi entropy $S_{n,\phi}$ for a highly excited state $|\phi\rag$, i.e., the conformal weight $h_\phi$ is order $c$ for short interval $\ell\ll L$ so that we can obtain the results with two different methods based on short interval expansion up to order $(\ell/L)^8$.

The first method is to use OPE of  twist operators on the cylinder to evaluate the excited state R\'enyi entropy \cite{Calabrese:2004eu,Headrick:2010zt,Calabrese:2010he,Chen:2013kpa}. We have used this method in  \cite{Lin:2016dxa} to get the result up to order $(\ell/L)^6$ and find that the subsystem ETH is violated for $n \neq 1$ but holds for $n=1$, i.e., the entanglement entropy.   In this paper we calculate up to order $(\ell/L)^8$ and find nontrivial violation of subsystem ETH at the new order. For consistency check we also use the other two methods to calculate and obtain the same result. The second method is to use the multi-point correlation functions on complex plane \cite{Alcaraz:2011tn,Berganza:2011mh,Lashkari:2014yva,Lashkari:2015dia,Sarosi:2016oks,Sarosi:2016atx,Ruggiero:2016khg}. As in \cite{Lin:2016dxa}, we focus on the contributions of the holomorphic sector of the vacuum conformal family. Some details of the vacuum conformal family are collected in appendix~\ref{vcf}.

\subsection{Method of twist operators}\label{sec3.1}

\begin{figure}[htbp]\centering
  \includegraphics[height=5.8cm]{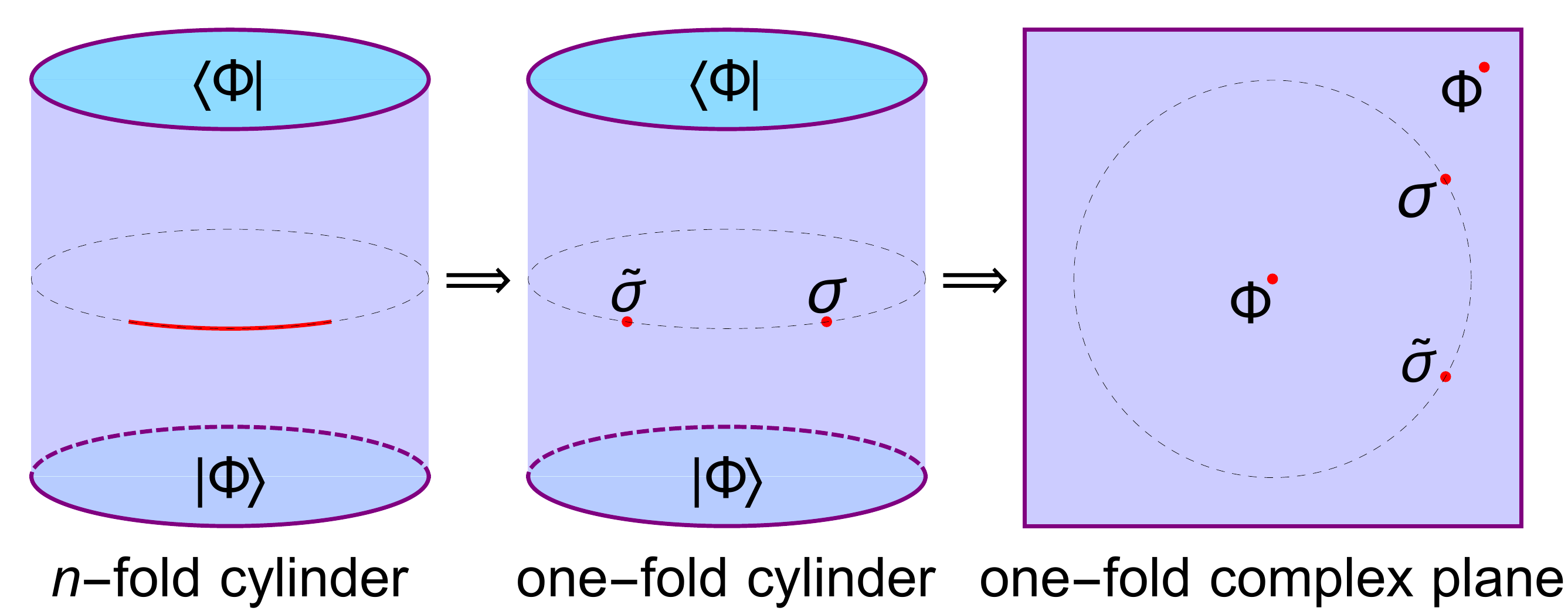}\\
  \caption{This figure illustrates how OPE of twist operators on a cylinder and that on a complex plane are related. The one-fold CFT on a one-fold CFT on an $n$-fold cylinder is equivalent to an $n$-fold CFT on a one-fold cylinder. The boundary conditions of the $n$-fold CFT can be replaced by the insertion of a pair of twist operators \cite{Calabrese:2004eu}. The cylinder with twist operators can be mapped to a complex plane with twist operators.}\label{RE12}
\end{figure}

By the replica trick for evaluating the single-interval R\'enyi entropy, we get the one-fold CFT on an $n$-fold cylinder, or equivalently an $n$-fold CFT, which we call $\CFT^n$ on the one-fold cylinder. The boundary conditions of the $\CFT^n$ on cylinder can be replaced by twist operators \cite{Calabrese:2004eu}. Thus, the partition function of $\CFT^n$ on cylinder capped by state $|\phi\rag$ can be expressed as the two-point function of twist operators, i.e.
\be
\tr_A\r_{A,\phi}^n = \lag\Phi|\s(\ell/2)\td\s(-\ell/2)|\Phi\rag_\cyl,
\ee
with the definition $\Phi \equiv \prod_{j=0}^{n-1}\phi_j$ and the index $j$ marking different replicas. This is illustrated in figure~\ref{RE12}.

Formally and practically, we can use the OPE of the twist operators to turn the above partition function into a series expansion, and the formal series expansion for the excited state R\'enyi entropy is
\be \label{e16}
S_{n,\phi}=\f{c(n+1)}{12n}\log\f{\ell}{\e} -\f{1}{n-1}\log \Big( \sum_K d_K \ell^{h_K} \lag\Phi_K\rag_\Phi \Big).
\ee
The details about the OPE of twist operators \cite{Headrick:2010zt,Calabrese:2010he,Chen:2013kpa} is reviewed in appendix~\ref{opests}. In arriving the above, we have used (\ref{sts}) and the fact that $\lag\Phi_K\rag_\Phi\equiv\lag\Phi|\Phi_K|\Phi\rag_\cyl$ is a constant.

  Further using the properties for the vacuum conformal family and its OPE in appendix~\ref{vcf} and  \ref{opests}, i.e.,  specifically (\ref{phikphi}), (\ref{PhiKPhi1}), (\ref{PhiKPhi2}) and (\ref{bK1}), we can obtain the explicit result of the short interval expansion up to order $(\ell/L)^8$ as follows:
\bea \label{Snphi8}
&& \hspace{-5mm}
S_{n,\phi} =  \frac{c(n+1)}{12n} \log \f{\ell}{\e}
        +\frac{\pi^2 c (n+1) (24 \epsilon_\phi-1)\ell^2}{72 n L^2}
        -\frac{\pi^4 c (n+1) \big[ 48 (n^2+11)24 \epsilon_{\phi}^2- 24(n^2+1)\epsilon_{\phi} +n^2 \big] \ell^4}{2160  n^3 L^4} \nn\\
&& \hspace{-5mm} \phantom{S_{n,\phi}=}
-\frac{\pi^6 c (n+1) \big[  96 (n^2-4) (n^2+47) \epsilon_\phi^3
                           +36 (2 n^4+9 n^2+37) \epsilon_\phi^2
                           -24 (n^4+n^2+1)\epsilon_\phi + n^4 \big] \ell^6}{34020  n^5 L^6} \nn\\
&& \hspace{-5mm} \phantom{S_{n,\phi}=}
+\frac{\pi^8 c (n+1) \ell^8}{453600 (5 c+22) n^7 L^8} \big\{ c \big[ 64 (13 n^6-1647 n^4+33927 n^2-58213) \e_\phi^4 \nn\\
&& \hspace{-5mm} \phantom{S_{n,\phi}=}                               -64(n^2+11) (13 n^4+160 n^2-533) \e_\phi^3
                                                                      -48 (9 n^6+29 n^4+71 n^2+251) \e_\phi^2  \nn\\
&& \hspace{-5mm} \phantom{S_{n,\phi}=}
                                                                      +120 (n^2+1) (n^4+1) \e_\phi
                                                                      -5 n^6 \big]
                                       -5632(n^2-4)(n^2-9)(n^2+119) \e_\phi^4 \nn\\
&& \hspace{-5mm} \phantom{S_{n,\phi}=}
                                       -2816 (n^2-4) (n^2+11) (n^2+19) \e_\phi^3
                                       -128 (15 n^6+50 n^4+134 n^2+539) \e_\phi^2\nn\\
&& \hspace{-5mm} \phantom{S_{n,\phi}=}
                                       +528 (n^2+1) (n^4+1) \e_\phi
                                       -22 n^6 \big\} +O((\ell/L)^{9}).
\eea
Note that the result up to order $(\ell/L)^6$ is just proportional to $c$, and agrees with the result obtained previously in \cite{Lashkari:2016vgj,Lin:2016dxa}. At the order $(\ell/L)^8$, however, novel property appears. There appears a nontrivial $5c+22$ factor in the overall denominator, which yields infinite number of higher order subleading terms in the $1/c$ expansion for large $c$. These subleading terms come from the contributions of the quasiprimary operator $\mA$ defined by \eq{Adef} at level four of the vacuum family.  We will obtain the same result for other method in  subsection~\ref{sec3.2}.

  Instead of working on the cylinder geometry, we can also work on complex plane by conformal map, as shown in figure~\ref{RE12}. The cylinder with coordinate $w$ is mapped to a complex plane with coordinate $z$ by a conformal transformation $z=\ep^{2\pi\ii w/L}$. The partition function then becomes a four-point function on complex plane
\be
\tr_A\r_{A,\phi}^n = \Big( \f{2\pi\ii}{L} \Big)^{2h_\s} \lag\Phi(\inf)\s(\ep^{\pi\ii\ell/L})\td\s(\ep^{-\pi\ii\ell/L})\Phi(0)\rag_\mC.
\ee
Using \eq{sts} for the OPE of twist operators on complex plane, we get the excited state R\'enyi entropy
\be
S_{n,\phi}=\f{c(n+1)}{12n}\log\Big(\f{L}{\pi\e}\sin\f{\pi\ell}{L}\Big)
          -\f1{n-1} \Big[ \sum_K d_K C_{\Phi\Phi K}(1-\ep^{2\pi\ii\ell/L})^{h_K}{_2}F_1(h_K,h_K;2h_K;1-\ep^{2\pi\ii\ell/L}) \Big].
\ee
Using (\ref{Cphiphik}), (\ref{CPhiPhiK1}), (\ref{CPhiPhiK2}), (\ref{CPhiPhiK3}), (\ref{CPhiPhiK4}), (\ref{bK1}), (\ref{bK2}) we can reproduce (\ref{Snphi8}).

\subsection{Method of multi-point function on complex plane}\label{sec3.2}

\begin{figure}[htbp]\centering
  \includegraphics[height=5.8cm]{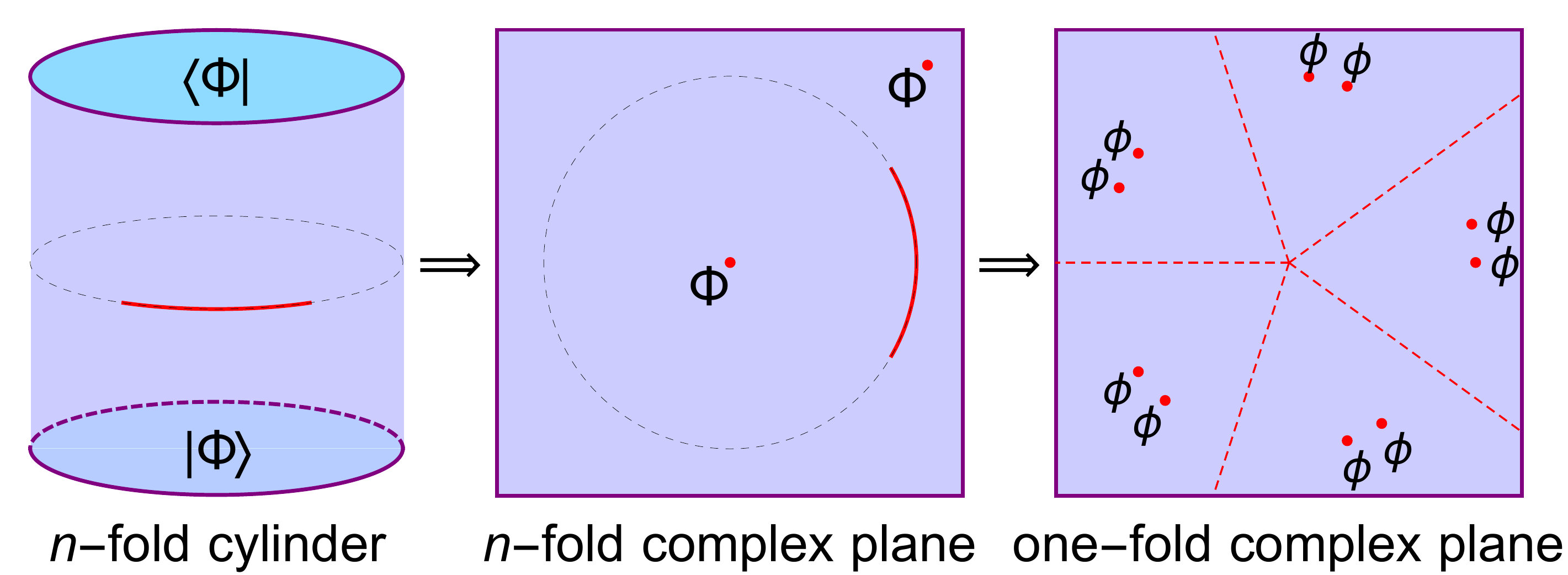}\\
  \caption{This figure illustrates the replica method of multi-point function on complex plane \cite{Alcaraz:2011tn,Berganza:2011mh,Lashkari:2014yva,Lashkari:2015dia,Sarosi:2016oks,Sarosi:2016atx,Ruggiero:2016khg}. Firstly, one has the one-fold CFT on an $n$-fold cylinder in excited state $|\Phi\rag$. Then, by state/operator correspondence one gets a two-point function on an $n$-fold complex plane $\mC^n$. Lastly, by a conformal transformation one gets a $2n$-point function on a one-fold complex plane. In the last part of the figure we use $n=5$ as an example.}\label{RE3}
\end{figure}

In the second method we use the formulism of multi-point function on complex plane, see \cite{Alcaraz:2011tn,Berganza:2011mh,Lashkari:2014yva,Lashkari:2015dia,Sarosi:2016oks,Sarosi:2016atx,Ruggiero:2016khg}. The idea is illustrated in Fig.\ref{RE3}. Using the state/operator correspondence, we map the partition function on the capped $n$-fold cylinder into the two-point function on the $n$-fold complex plane $\mC^n$, i.e., formally
\be
\f{\tr_A \r_{A,\phi}^n}{\tr_A \r_{A,0}^n} = \lag\Phi(\inf)\Phi(0)\rag_{\mC^n}.
\ee
We then map each copy of complex plan into a wedge of deficit angle $2\pi/n$ by the following  conformal transformation
\be
f(z) = \Big( \f{z-\ep^{\pi\ii\ell/L}}{z-\ep^{-\pi\ii\ell/L}} \Big)^{1/n}.
\ee
The two boundaries of each wedge correspond to the intervals just right above or below the interval $A$. Gluing all the $n$ wedges along the boundaries, we then obtain the one-fold complex plane $\mC$ so that the above two-point function on $\mC^n$ becomes a $2n$-point function on a one-fold complex plane $\mC$, i.e.
\be \label{e28}
\lag\Phi(\inf)\Phi(0)\rag_{\mC^n} = \Big( \f{2\ii}{n}\sin\f{\pi\ell}{L} \Big)^{2nh_\phi}
   \Big\lag \prod_{j=0}^{n-1} \Big( \ep^{2\pi\ii(\f{\ell}{n L}+\f{2j}{n})}
                                    \phi(\ep^{2\pi\ii(\f{\ell}{n L}+\f{j}{n})})
                                    \phi(\ep^{2\pi\ii \f{j}{n}}) \Big)  \Big\rag_\mC.
\ee
Based on the above, it is straightforward to see that
\be
S_{n,\phi}(\ell) = S_{n,\phi}(L-\ell),
\ee
which is expected for a pure state.

 Formally, the OPE of a primary operator with itself is given \cite{Blumenhagen:2009zz}
\be \label{phizphiw}
\phi(z)\phi(w) = \f{1}{(z-w)^{2h_\phi}}\mF_\phi(z,w).
\ee
with
\be
\mF_\phi(z,w)=\sum_\mX \f{C_{\phi\phi\mX}}{\a_\mX} \sum_{r=0}^\inf \f{a_\mX^r}{r!}(z-w)^{h_\mX+r}\p^r\mX(w), ~~
a_\mX^r = \f{C_{h_\mX+r-1}^r}{C_{2h_\mX+r-1}^r},
\ee
where the summation runs over all the holomorphic quasiprimary operators $\{ \mX \}$ with each $\mX$ being of conformal weight $h_\mX$, and $C_x^y$ denotes the binomial coefficient.

In a unitary CFT, the operator with the lowest conformal weight is the identity operator, and so in $z\to w$ limit
\be
\mF_\phi(z,w) = 1 + \cdots.
\ee
Putting (\ref{phizphiw}) in (\ref{e28}) and using \cite{Calabrese:2004eu}
\be
\tr_A \r_{A,0}^n = \Big(\f{L}{\pi\e}\sin\f{\pi\ell}{L}\Big)^{-2h_\s},
\ee
we get the excited state R\'enyi entropy
\be \label{Snphi3}
S_{n,\phi} =  \f{c(n+1)}{12n}\log\Big( \f{L}{\pi\e}\sin\f{\pi\ell}{L} \Big)
             -\f{2nh_\phi}{n-1}\log\f{\sin\f{\pi\ell}{L}}{n\sin\f{\pi\ell}{nL}}
             -\f{1}{n-1}\log \Big\lag \prod_{j=0}^{n-1}\mF_\phi\big( \ep^{2\pi\ii(\f{\ell}{n L}+\f{j}{n})},
                                                                     \ep^{2\pi\ii \f{j}{n}} \big) \Big\rag_\mC.
\ee

We now perform the short-interval expansion for \eq{Snphi3} up to order $(\ell/L)^8$ by considering only the contributions from the vacuum conformal family, i.e.,  including its descendants up to level eight, see appendix~\ref{vcf}. For $n=2$ there is a compact formula
\be
S_{2,\phi} =  \f{c}{8}\log\Big( \f{L}{\pi\e}\sin\f{\pi\ell}{L} \Big)
             -4h_\phi\log\f{\sin\f{\pi\ell}{L}}{2\sin\f{\pi\ell}{2L}}
             -\log \Big[  \sum_\psi \f{C_{\phi\phi\psi}^2}{\a_\psi} \Big(\sin\f{\pi\ell}{2L}\Big)^{2h_\psi}
                          {}_2F_1(h_\psi,h_\psi;2h_\psi;\sin^2\f{\pi\ell}{2L})  \Big].
\ee
Using the details in the appendix~\ref{vcf}, we can obtain the explicit result as follows\footnote{Since  $C_{\phi\phi\mX}=0$ for a bosonic operator $\mX$ with odd integer conformal weight, using results in the appendix~\ref{vcf} we can get the result up to order $(\ell/L)^{19}$. However higher order results are too complicated to be revealing. We just write down the result up to order $(\ell/L)^{11}$.}
\bea
&& \hspace{-8mm}
   S_{2,\phi} = \frac{c}{8} \log\f{\ell}{\e}
            +\frac{\pi^2 c (24 \e_\phi -1)\ell^2}{48 L^2}
            -\frac{\pi^4 c (180 \e_\phi^2-30 \e_\phi+1)\ell^4}{1440 L^4}
            -\frac{\pi^6 c (945 \e_\phi^2-126 \e_\phi+4)\ell^6}{90720 L^6}  \nn\\
&& \hspace{-8mm}\phantom{S_{2,\phi} =}
            +\frac{\pi^8 c [ 5c(83160 \e_\phi^4-7560 \e_\phi^3-1890 \e_\phi^2+255 \e_\phi-8)
                            -2 (22680 \e_\phi^2-2805 \e_\phi+88) ] \ell^8}{2419200 (5 c+22) L^8}  \nn\\
&& \hspace{-8mm}\phantom{S_{2,\phi} =}
            +\frac{\pi^{10} c [5 c (1372140 \e_\phi^4-124740 \e_\phi^3-14355 \e_\phi^2+2046 \e_\phi-64)
                                        -44 (8595 \e_\phi^2-1023 \e_\phi+32)]\ell^{10}}{239500800 (5 c+22) L^{10}}\nn\\
&& \hspace{-8mm}\phantom{S_{2,\phi} =}
            +O((\ell/L)^{12}).
\eea

To perform the short-interval expansion R\'enyi entropy of general rank $n$, i.e., (\ref{Snphi3}), to order $\ell^{m}$, we have to calculate a series of $j$-point correlation functions with $j=1,2,\cdots,\lfloor m/2 \rfloor$. To order $\ell^8$ the number of these multi-point correlation functions can be counted in each order by the following:
\bea \label{counting}
&& \prod_{k=2}^\inf \f{1}{(1-x^k)^n} = 1+n x^2+n x^3+\frac{n (n+3)}{2} x^4 +n (n+1) x^5+\frac{n (n+1) (n+11)}{6} x^6 \nn\\
&& \phantom{\prod_{k=2}^\inf \f{1}{(1+x^k)^n} =}
   +\frac{n (n^2+5n+2)}{2} x^7+\frac{n (n+3) (n^2+27n+14)}{24} x^8+O(x^9).
\eea
These multi-point correlation functions are listed in table~\ref{jptfn}, and we note that many of them are trivially vanishing. Putting the  results of appendix~\ref{vcf} in (\ref{Snphi3}), we reproduce the R\'enyi entropy (\ref{Snphi8}).

\begin{table}[htbp]\centering
\begin{tabular}{|c|c|l|c|c|}\cline{1-5}
  order & multi-point function & ?~?~?~?~? & number & number \\ \cline{1-5}

  0                & 1                    & \checked  \texttimes\texttimes\texttimes\texttimes & 1 & 1 \\ \cline{1-5}

  2                & $T$                  & \texttimes\checked  \texttimes\texttimes\texttimes  & $n$ & $n$ \\ \cline{1-5}

  3                & $\p T$               & \texttimes\checked  \texttimes & $n$ & $n$ \\ \cline{1-5}

                   & $\mA$                & \texttimes\texttimes \checked  & $n$             &                     \\ \cline{2-4}
  4                & $TT$                 & \checked  \checked  \checked   & $\f{n(n-1)}{2}$ & $\frac{n (n+3)}{2}$ \\ \cline{2-4}
                   & $\p^2T$              & \texttimes\checked  \texttimes & $n$             &                     \\ \cline{1-5}

                   & $\p^3T$              & \texttimes\checked   & $n$      &          \\ \cline{2-4}
  5                & $\p\mA$              & \texttimes\texttimes & $n$      & $n(n+1)$ \\\cline{2-4}
                   & $T\p T$              & \checked  \checked   & $n(n-1)$ &          \\ \cline{1-5}

                   & $\mB$, $\mD$         & \texttimes\texttimes & $2n$                 &                            \\ \cline{2-4}
                   & $T\mA$               & \texttimes\checked   & $n(n-1)$             &                            \\ \cline{2-4}
  \multirow{2}*{6} & $TTT$                & \checked  \checked   & $\f{n(n-1)(n-2)}{6}$ & \multirow{2}*{$\frac{n (n+1) (n+11)}{6}$} \\ \cline{2-4}
                   & $\p^4T$              & \texttimes\checked   & $n$                  &                            \\ \cline{2-4}
                   & $\p^2\mA$            & \texttimes\texttimes & $n$                  &                            \\ \cline{2-4}
                   & $T\p^2T$, $\p T\p T$ & \checked  \checked   & $\f{3n(n-1)}{2}$     &                            \\ \cline{1-5}

                   & $\p^5T$, $\p^3\mA$, $\p\mB$, $\p\mD$   & \texttimes & $4n$                 &                                         \\ \cline{2-4}
  \multirow{2}*{7} & $T\p^3T$, $\p T\p^2 T$                 & \checked   & $2n(n-1)$            & \multirow{2}*{$\frac{n (n^2+5n+2)}{2}$} \\ \cline{2-4}
                   & $T\p\mA$, $\p T\mA$                    & \texttimes & $2n(n-1)$            &                                         \\ \cline{2-4}
                   & $TT\p T$                               & \checked   & $\f{n(n-1)(n-2)}{2}$ &                                         \\ \cline{1-5}

    & $\mE$, $\mH$, $\mI$                      & \texttimes & $3n$                       & \\ \cline{2-4}
    & $T\mB$, $T\mD$                           & \texttimes & $2n(n-1)$                  & \\ \cline{2-4}
    & $\mA\mA$                                 & \checked   & $\f{n(n-1)}{2}$            & \\ \cline{2-4}
    & $TT\mA$                                  & \checked   & $\f{n(n-1)(n-2)}{2}$       & \\ \cline{2-4}
  8 & $TTTT$                                   & \checked   & $\f{n(n-1)(n-2)(n-4)}{24}$ & $\frac{n (n+3) (n^2+27n+14)}{24}$ \\ \cline{2-4}
    & $\p^6T$, $\p^4\mA$, $\p^2\mB$, $\p^2\mD$ & \texttimes & $4n$                       & \\ \cline{2-4}
    & $T\p^4T$, $\p T\p^3T$, $\p^2T\p^2T$      & \checked   & $\f{5n(n-1)}{2}$           & \\ \cline{2-4}
    & $T\p^2\mA$, $\p T\p\mA$, $\p^2T\mA$      & \texttimes & $3n(n-1)$                  & \\ \cline{2-4}
    & $TT\p^2T$, $T\p T\p T$                   & \checked   & $n(n-1)(n-2)$              & \\ \cline{1-5}
\end{tabular}
\caption{All the multi-point functions we have to consider in obtaining the excited state R\'enyi entropy up to order $(\ell/L)^8$.
In the 1st column it is the order from which each multi-point function starts to contribute.
In the 2nd column listed are the multi-point functions, and for simplicity we have omitted the correlation function symbol $\lag \cdots \rag_\mC$ and the positions of the operators, which should be $\ep^{2\pi\ii j_1/n}, \ep^{2\pi\ii j_2/n}, \cdots$. The $j$'s take values from $0,1,\cdots,n-1$ with constraints that can be figured out easily. For example, $TT\mA$ denotes the three-point functions on complex plane $\lag T(\ep^{2\pi\ii j_1/n})T(\ep^{2\pi\ii j_2/n})\mA(\ep^{2\pi\ii j_3/n}) \rag_\mC$ with $0\leq j_{1,2,3}\leq n-1$ and the constraints $j_1<j_2, j_1 \neq j_3, j_2\neq j_3$. In the 3rd column we mark the answers to several questions. For the 1st question we mark \checked{} if the multi-point functions are non-vanishing and we \texttimes{} if they are vanishing. The 2nd, 3rd, 4th and 5th questions are for the calculation of (\ref{ee36}) in section~\ref{sec5}, and they are about whether the multi-point functions are non-vanishing or vanishing after the insertion, respectively, of the operators $T$, $\mA$, $\mB$ and $\mD$. In the 4th and 5th columns are the numbers of the multi-point functions as counted by (\ref{counting}). Note that the table is similar to but different from table~\ref{cftnqp} which counts the $\CFT^n$ quasiprimary operators.}
\label{jptfn}
\end{table}

\section{Check subsystem weak ETH}\label{sec4}

We now can use the result in the previous section to check the subsystem weak ETH for the 2D large $c$ CFT. We first take the $n\to 1$ limit of the excited state R\'enyi entropy \eq{Snphi8} and the thermal one \eq{Snb} up to order $(\ell/L)^8$ to get the corresponding entanglement entropy. We then have
\bea \label{Sphi8}
&& \hspace{-7mm}
  S_{A,\phi} = \frac{c}{6} \log \f{\ell}{\e}
   +\frac{\pi^2c (24 \epsilon_\phi-1)\ell^2}{36 L^2}
   -\frac{\pi^4c (24 \epsilon_\phi-1)^2\ell^4}{1080 L^4}
   +\frac{\pi^6c (24 \epsilon_\phi-1)^3\ell^6}{17010 L^6}
   -\frac{\pi^8 c \ell^8}{226800 (5 c+22) L^8}
   \big[ 5c (24 \e_\phi-1)^4 \nn\\
&& \hspace{-7mm}\phantom{S_{A,\phi} =}
        +2 (8110080 \e_\phi^4
        -1013760 \e_\phi^3
        +47232 \e_\phi^2
        -1056 \e_\phi
        +11) \big]
   +O((\ell/L)^{9}),
\eea
and
\be \label{Sbeta}
S_{A,\b} = \frac{c}{6} \log\f{\ell}{\e}+\frac{\pi^2 c \ell^2}{36 \beta^2}-\frac{\pi^4 c \ell^4}{1080 \beta^4}+\frac{\pi^6 c \ell^6}{17010 \beta^6}-\frac{\pi^8 c \ell^8}{226800 \beta^8} +O((\ell/\b)^{10}).
\ee
It is straightforward to see that (\ref{Sphi8}) fails to match with (\ref{Sbeta}) at order $(\ell/L)^8$ under the identification of inverse temperature and conformal weight by the relation \eq{e47}, and the discrepancy is
\be\label{DS-1}
 S_{A,\b} - S_{A,\phi} = \frac{128\pi^8c\e_\phi^2(22\e_\phi-1)^2\ell^8}{1575(5 c+22) L^8}+O((\ell/L)^{9}).
\ee
From \eq{LDL} we know that this is nothing but the relative entropy $S(\r_{A,\phi}\|\r_{A,\b})$. Note that this discrepancy is of order $c^0$ in the large $c$ limit, and also there are infinite number of subleading terms in large $c$ expansion.

Based on the result \eq{DS-1}, we then use the inequalities \eq{FAI}, \eq{AI}, and \eq{PI} to estimate the order of the trace distance in large $c$ limit and check the validity of ETH. We have obtained the R\'enyi entropy, entanglement entropy and relative entropy firstly in expansion of small $\ell$, and then in expansion of large $c$. Focusing on the order of large $c$, we have
\bea
&& \D S_A \sim \mO(c^0), \nn\\
&& \D S_n \sim \mO(c), \nn\\
&& S(\r_{A,\phi}\|\r_{A,\b}) \sim \mO(c^0),
\eea
and we assume that these orders still apply when $0<\ell/L<1$ is neither too small nor too large.  From \eq{FAI}, \eq{AI}, and \eq{PI} we get respectively
\bea \label{z58}
&& t \geq \f{\mO(c^0)}{\log d}, \nn\\
&& d \geq \ep^{\mO(c)}, \nn\\
&& t \leq \mO(c^0).
\eea
From the first inequality of \eq{z58}, the lower bound of the trace distance $t$, which is crucial for the validity of subsystem weak ETH, depends on if the effective dimension $d$ of the subsystem $A$ is strictly infinite or how it scales with $c$. It is a subtle issue to determine $d$ for generic CFTs. We will raise this as an interesting issue for further study, but now consider some interesting scenarios.\footnote{In \cite{Dymarsky:2016aqv}, instead a similar proposal to our first inequality of \eq{z58} for subsystem weak ETH, i.e.,  $t\sim d \; \ep^{-{1\over 2} S(E)}$ is used as the definition for the effective dimension $d$. We understand this as the working definition because the spectrum of the density matrix for the CFT should be continuous without further coarse-graining.}

If $d$ is strictly infinite, from (\ref{z58}) we get
\be\label{check-1}
0 \leq t \leq \mO(c^0),
\ee
which is trivial and gives no useful information.
It is also possible there exists a large but finite effective dimension $d$ of the subsystem $A$  that satisfies \eq{FAI}, \eq{AI} and so satisfies \eq{z58}. The Cardy's formula \cite{Cardy:1986ie} and Boltzmann's entropy formula $\Omega(E) \sim \ep^{S(E)}$ state that the number of states at a specific high energy is $\ep^{\mO(c)}$. It is plausible that for both the reduced density matrices $\r_{A,\phi}$ and $\r_{A,\b}$ only $\ep^{\mO(c)}$ components are nontrivial and other components are even smaller than exponential suppression. Then we get the tentative result
\be
d \sim \ep^{\mO(c)}.
\ee
If this is true, from (\ref{z58}) we get
\be\label{check-2}
\mO (c^{-1}) \leq t \leq \mO(c^0).
\ee

Both \eq{check-1} and \eq{check-2} are consistent with the subsystem weak ETH \eq{weakETH}. However, it lacks further evidence to obtain the power of suppression.


\section{Relative entropy between primary states}\label{sec5}

In this section we present some byproducts of this paper obtained by using the same method as the one in subsection~\ref{sec3.2} \cite{Alcaraz:2011tn,Berganza:2011mh,Lashkari:2014yva,Lashkari:2015dia,Sarosi:2016oks,Sarosi:2016atx,Ruggiero:2016khg}, which has also been used to calculate the relative entropy \cite{Lashkari:2014yva,Sarosi:2016oks,Sarosi:2016atx,Ruggiero:2016khg}. We will calculate the relative entropy $S(\r_{A,\phi}\|\r_{A,\psi})$, the 2nd symmetrized relative entropy $S_2(\r_{A,\phi},\r_{A,\psi})$, and the Schatten 2-norm $\|\r_{A,\phi}-\r_{A,\psi} \|_2$ between the reduced density matrices of two primary states $|\phi\rag$ and $|\psi\rag$ in the short interval expansion, where $\psi$ is similar to $\phi$ and is the primary field of conformal weight $h_\psi=c\e_\psi$.

To calculate the relative entropy $S(\r_{A,\phi}\|\r_{A,\psi})$, we first need to  calculate the ``$n$-th relative entropy''
\be
S_n(\r_{A,\phi}\|\r_{A,\psi}) = \f{1}{n-1}  \big( \log \tr_A \r_{A,\phi}^n -  \log\tr_A (\r_{A,\phi}\r_{A,0}^{n-1}) \big).
\ee
and then take $n\to 1$ limit.

\begin{figure}[htbp]\centering
  \includegraphics[height=5.8cm]{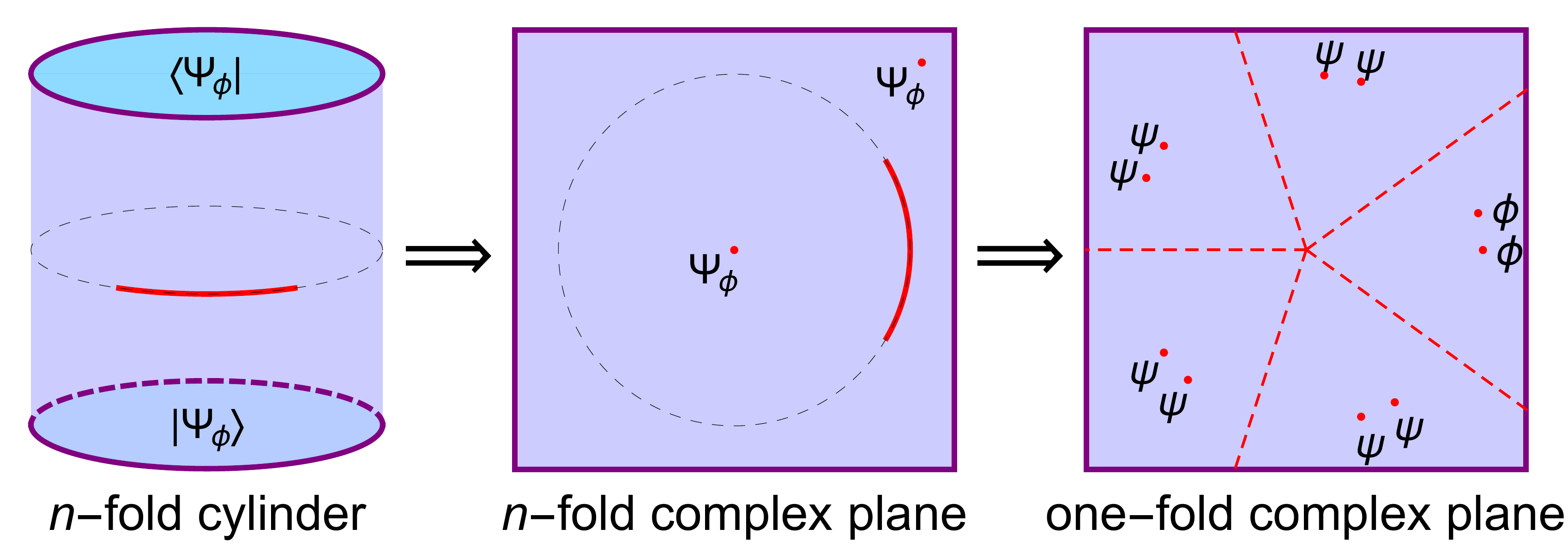}\\
  \caption{The calculation of $\tr_A (\r_{A,\phi}\r_{A,\psi}^{n-1})$. Here $\Psi_\phi\equiv\phi_0\prod_{j=1}^{n-1}\psi_j$, with $\phi$ existing in one copy and $\psi$ existing in the other $n-1$ copies.}\label{relative}
\end{figure}

We have already calculated $\tr_A \r_{A,\phi}^n$ as illustrated in figure~\ref{RE3} of in subsection~\ref{sec3.2}, and this inspires us to calculate $\tr_A (\r_{A,\phi}\r_{A,\psi}^{n-1})$ as illustrated in figure~\ref{relative}.  Similar to the manipulation in subsection~\ref{sec3.2}, in the end we can obtain the formal result
\be \label{ee36}
\f{\tr_A (\r_{A,\phi}\r_{A,\psi}^{n-1})}{\tr_A \r_{A,0}^n}
=\lag \Psi_\phi(\inf)\Psi_\phi(0) \rag_{\mC^n}
= \Big( \f{\sin\f{\pi\ell}{L}}{n\sin\f{\pi\ell}{nL}} \Big)^{2(h_\phi+(n-1)h_\psi)}
\Big\lag
\mF_\phi\big( \ep^{2\pi\ii\f{\ell}{n L}},1 \big)
\prod_{j=1}^{n-1}
\mF_\psi\big( \ep^{2\pi\ii(\f{\ell}{n L}+\f{j}{n})},\ep^{2\pi\ii \f{j}{n}} \big)
\Big\rag_\mC,
\ee
Here $\Psi_\phi\equiv\phi_0\prod_{j=1}^{n-1}\psi_j$, with $\phi$ existing in one copy and $\psi$ existing in the other $n-1$ copies.
The explicit result up to order $(\ell/L)^8$ is
\bea
&& S_n(\r_{A,\phi}\|\r_{A,\psi}) = 2c(\e_\phi-\e_\psi)\log\f{\sin\f{\pi\ell}{L}}{n\sin\f{\pi\ell}{nL}}
                                  +\frac{\pi^4 c(\e_\phi-\e_\psi) (n+1) (n^2+11) \ell^4}
                                        {45 n^4 L^4}
                                   \big(n \e_\phi+(n-2) \e_{\psi}\big) \nn\\
&& \phantom{S_n(\r_{A,\phi}\|\r_{A,\psi}) =}
  +\frac{\pi^6 c(\e_\phi-\e_{\psi})(n+1) \ell^6}{2835 L^6 n^6}
   \big( 8 n (n^2-4)(n^2+47) \e_\phi^2
        +8 (n-3)(n^2-4) (n^2+47) \e_\psi^2  \nn\\
&& \phantom{S_n(\r_{A,\phi}\|\r_{A,\psi}) =}
        +8 n (n^2-4)(n^2+47) \e_\phi\e_\psi
        +3 n (2 n^4+9 n^2+37) \e_\phi
        +3 (n-2) (2 n^4+9 n^2+37) \e_{\psi} \big) \nn\\
&& \phantom{S_n(\r_{A,\phi}\|\r_{A,\psi}) =}
  -\frac{\pi^8 c(\e_\phi-\e_{\psi}) (n+1)\ell^8}{28350 (5 c+22) n^8 L^8}
   \big[c \big( 4 n (13 n^6-1647 n^4+33927n^2-58213) \e_\phi^3 \nn\\
&& \phantom{S_n(\r_{A,\phi}\|\r_{A,\psi}) =}
           +4 (n-2) (13 n^6+40 n^5-1567 n^4+4400 n^3+42727n^2-42840 n-143893) \e_\psi^3 \nn\\
&& \phantom{S_n(\r_{A,\phi}\|\r_{A,\psi}) =}
           +4 n (13 n^6-1647 n^4+33927n^2-58213) \e_\phi^2\e_\psi \nn\\
&& \phantom{S_n(\r_{A,\phi}\|\r_{A,\psi}) =}
           +4 (n-2) (13 n^6-40 n^5-1727 n^4-4400n^3+25127 n^2+42840 n+27467) \e_\phi\e_\psi^2 \nn\\
&& \phantom{S_n(\r_{A,\phi}\|\r_{A,\psi}) =}
           -4 n(n^2+11) (13 n^4+160 n^2-533) \e_\phi^2 \\
&& \phantom{S_n(\r_{A,\phi}\|\r_{A,\psi}) =}
           -4 (n-2) (n^2+11) (13 n^4-10 n^3+140 n^2-190 n-913) \e_\psi^2 \nn\\
&& \phantom{S_n(\r_{A,\phi}\|\r_{A,\psi}) =}
           -4 (n^2+11) (13 n^5-3 n^4+160 n^3-10 n^2-533 n-227)\e_\phi\e_\psi \nn\\
&& \phantom{S_n(\r_{A,\phi}\|\r_{A,\psi}) =}
           -3 n (9 n^6+29 n^4+71 n^2+251) \e_\phi
           -3 (n-2) (9n^6+29 n^4+71 n^2+251) \e_\psi \big)  \nn\\
&& \phantom{S_n(\r_{A,\phi}\|\r_{A,\psi}) =}
    -352 n (n^2-9) (n^2-4)(n^2+119) \e_\phi^3
    -352 (n-4)(n^2-9) (n^2-4) (n^2+119) \e_\psi^3  \nn\\
&& \phantom{S_n(\r_{A,\phi}\|\r_{A,\psi}) =}
    -352 n (n^2-9) (n^2-4) (n^2+119) \e_\phi^2\e_\psi
    -352 n (n^2-9) (n^2-4) (n^2+119) \e_\phi\e_{\psi}^2  \nn\\
&& \phantom{S_n(\r_{A,\phi}\|\r_{A,\psi}) =}
    -176 n (n^2-4) (n^2+11)(n^2+19) \e_\phi^2
    -176 (n-3)(n^2-4) (n^2+11) (n^2+19) \e_\psi^2  \nn\\
&& \phantom{S_n(\r_{A,\phi}\|\r_{A,\psi}) =}
    -176 n (n^2-4) (n^2+11) (n^2+19) \e_\phi\e_\psi
    -8 n (15 n^6+50 n^4+134 n^2+539) \e_{\phi}  \nn\\
&& \phantom{S_n(\r_{A,\phi}\|\r_{A,\psi}) =}
    -8 (n-2) (15 n^6+50 n^4+134 n^2+539)\e_\psi \big] + O((\ell/L)^9). \nn
\eea
By taking $n\to 1$ limit we get
\bea\label{e515}
&& S(\r_{A,\phi}\|\r_{A,\psi}) =
   \frac{8 \pi^4 c (\e_\phi-\e_\psi)^2\ell^4}{15L^4}
  -\frac{32\pi^6 c (\e_\phi-\e_\psi)^2\big(8 (\e_\phi+2 \e_\psi)-1\big)\ell^6}{315 L^6} \nn\\
&& \phantom{S(\r_{A,\phi}\|\r_{A,\psi}) =}
  +\frac{8 \pi^8 c (\e_\phi-\e_\psi)^2\ell^8}{1575 (5 c+22) L^8}
    \big(5 c (288 \e_\phi^2+1568 \e_\psi^2+576\e_\phi \e_\psi-48 \e_\phi-128 \e_\psi+3) \\
&& \phantom{S(\r_{A,\phi}\|\r_{A,\psi}) =}
          +2 (7040 \e_\phi^2+21120 \e_\psi^2+14080 \e_\phi\e_\psi-880 \e_\phi-1760 \e_\psi+41) \big)
+ O((\ell/L)^9). \nn
\eea
To order $\ell^6$ the result is in accord with \cite{Sarosi:2016oks,Sarosi:2016atx}.

Using the above result we can obtain $S(\r_{A,\psi}\|\r_{A,\phi})$ by swapping $\e_\phi$ and $\e_\psi$ in \eq{e515}. After that we can get the symmetrized relative entropy
\bea
&& S(\r_{A,\phi},\r_{A,\psi}) =
  \frac{16 \pi^4 c (\e_\phi-\e_\psi)^2\ell^4}{15L^4}
  -\frac{64 \pi^6 c (\e_\phi-\e_\psi)^2(12 (\e_\phi+\e_\psi)-1) \ell^6}{315L^6} \nn\\
&& \phantom{ d(\r_{A,\phi},\r_{A,\psi}) =}
  +\frac{16 \pi^8 c(\e_\phi-\e_\psi)^2 \ell^8}{1575(5 c+22) L^8}
  \big[5c ( 928 (\e_\phi^2+\e_\psi^2)+576 \e_\phi \e_\psi-88 (\e_\phi+\e_{\psi})+3)\nn\\
&& \phantom{ d(\r_{A,\phi},\r_{A,\psi}) =}
   +2 \big(14080 (\e_\phi^2+\e_\psi^2)+14080\e_\phi \e_\psi-1320 (\e_\phi+\e_{\psi})+41\big)\big]
+ O((\ell/L)^9).
\eea
Note that if we take $\e_\psi=0$, we can obtain $S(\r_{A,\phi}\|\r_{A,0})$, $S(\r_{A,0}\|\r_{A,\phi})$ and $S(\r_{A,\phi},\r_{A,0})$ which characterize the difference of the excited state $|\phi\rag$ and the vacuum state $|0\rag$. Moreover, all the above results show nontrivial subleading $1/c$ corrections at the order $(\ell/L)^8$.

The $n$-th relative entropy for $n\ne 1$ is not positive definite so that it cannot used as the measure for the difference of two quantum states. However, it turns out the 2nd symmetrized relative entropy $S_2(\r,\r')$ is positive definite because it can be written as
\be
S_2(\r,\r') = \log \f{\tr\r^2 \tr\r'^2}{[\tr(\r\r')]^2}.
\ee
Thus, $S_2(\r,\r')$ can be used as a difference measure between two quantum states. In fact it is directly related to the overlap of the two density matrices
\be
\mF(\r,\r') = \f{[\tr(\r\r')]^2}{\tr\r^2 \tr\r'^2}.
\ee
Note that the 2nd symmetrized relative entropy $S_2(\r,\r')$ is vanishing if and only if two density matrices are identical $\r=\r'$ and is infinite for two orthogonal density matrices $\tr(\r\r') =0$.

More general, one can also use Schatten $n$-norm to measure the difference of two density matrices
\be
\| \r-\r' \|_n = [\tr ( |\r-\r'|^n) ]^{1/n}.
\ee
For $n=1$ it is just the trace distance, and for $n=2$ we have
\be
\| \r-\r' \|_2 = [ \tr\r^2 + \tr\r'^2 -2 \tr(\r\r') ]^{1/2},
\ee

Below we will calculate both $S_2(\r_{A,\phi},\r_{A,\psi})$ and $\| \r_{A,\phi}-\r_{A,\psi} \|_2$ by following the similar trick used in the previous subsection. In fact, the ingredients needed to carry out the calculations such as $\tr_A\r_{A,\phi}^2$ and $\tr_A(\r_{A,\phi}\r_{A,\psi})$ have all been done already in previous sections. Packing them up, we then obtain the formal results as follows:
\bea
&& S_2(\r_{A,\phi},\r_{A,\psi}) = \log \bigg\{ \Big[  \sum_\mX \f{C_{\phi\phi\mX}^2}{\a_\mX} \Big(\sin\f{\pi\ell}{2L}\Big)^{2h_\mX}
                                                       {}_2F_1(h_\mX,h_\mX;2h_\mX;\sin^2\f{\pi\ell}{2L})  \Big] \nn\\
&& \phantom{S_2(\r_{A,\phi},\r_{A,\psi}) =}
                                       \times\Big[  \sum_\mX \f{C_{\psi\psi\mX}^2}{\a_\mX} \Big(\sin\f{\pi\ell}{2L}\Big)^{2h_\mX}
                                                       {}_2F_1(h_\mX,h_\mX;2h_\mX;\sin^2\f{\pi\ell}{2L})  \Big] \nn\\
&& \phantom{S_2(\r_{A,\phi},\r_{A,\psi}) =}
                                       \div\Big[  \sum_\mX \f{C_{\phi\phi\mX}C_{\psi\psi\mX}}{\a_\mX} \Big(\sin\f{\pi\ell}{2L}\Big)^{2h_\mX}
                                                       {}_2F_1(h_\mX,h_\mX;2h_\mX;\sin^2\f{\pi\ell}{2L})  \Big]^2\bigg\},  
\eea\bea
&&\hspace{-3mm} \|\r_{A,\phi}-\r_{A,\psi} \|_2 = \Big( \f{L}{\pi\e} \sin \f{\pi\ell}{L} \Big)^{-c/16}
\bigg\{ \Big( \f{\sin\f{\pi\ell}{L}}{2\sin\f{\pi\ell}{2L}} \Big)^{4h_\phi}
                                         \Big[  \sum_\mX \f{C_{\phi\phi\mX}^2}{\a_\mX} \Big(\sin\f{\pi\ell}{2L}\Big)^{2h_\mX}
                                                          {}_2F_1(h_\mX,h_\mX;2h_\mX;\sin^2\f{\pi\ell}{2L})  \Big] \nn\\
&&\hspace{-3mm} \phantom{\|\r_{A,\phi}-\r_{A,\psi} \|_2 =}
       +\Big( \f{\sin\f{\pi\ell}{L}}{2\sin\f{\pi\ell}{2L}} \Big)^{4h_\psi}
                                         \Big[  \sum_\mX \f{C_{\psi\psi\mX}^2}{\a_\mX} \Big(\sin\f{\pi\ell}{2L}\Big)^{2h_\mX}
                                                          {}_2F_1(h_\mX,h_\mX;2h_\mX;\sin^2\f{\pi\ell}{2L})  \Big]\\
&&\hspace{-3mm} \phantom{\|\r_{A,\phi}-\r_{A,\psi} \|_2 =}
       -2\Big( \f{\sin\f{\pi\ell}{L}}{2\sin\f{\pi\ell}{2L}} \Big)^{2(h_\phi+h_\psi)}
                                         \Big[  \sum_\mX \f{C_{\phi\phi\mX}C_{\psi\psi\mX}}{\a_\mX} \Big(\sin\f{\pi\ell}{2L}\Big)^{2h_\mX}
                                                          {}_2F_1(h_\mX,h_\mX;2h_\mX;\sin^2\f{\pi\ell}{2L})  \Big] \bigg\}^{1/2}.\nn
\eea
Then, in the short interval expansion we obtain the explicit results as follows:
\bea
&& \hspace{-3mm} S_2(\r_{A,\phi},\r_{A,\psi}) = \frac{\pi^4 c (\e_\phi-\e_\psi)^2\ell^4}{8 L^4}
 \Big[ 1
 +\frac{\pi^2 \ell^2}{12 L^2}
 +\frac{\pi^4 (5 c+24-20 c(\epsilon_{\phi}+\epsilon_{\psi}) (11 \epsilon_{\phi}+11 \epsilon_{\psi}-1))\ell^4}{160 (5 c+22)L^4} \nn\\
&& \hspace{-3mm}\phantom{S_2(\r_{A,\phi},\r_{A,\psi}) =}
 +\frac{\pi^6 (145 c+764-1260 c (\epsilon_{\phi}+\epsilon_{\psi}) (11 \epsilon_{\phi}+11 \epsilon_{\psi}-1))\ell^6}{60480 (5 c+22)L^6}
 +O((\ell/L)^8) \Big],\\
&& \hspace{-3mm}\|\r_{A,\phi}-\r_{A,\psi} \|_2 = \f{\pi^2 \sqrt{c(c+2)}(\e_\phi-\e_\psi)^2\ell^2}{4L^2}\Big( \f{\ell}{\e} \Big)^{-c/16}
 \Big[ 1
 +\frac{\pi^2 (c+4) (c+2-12 c (\epsilon_{\phi}+\epsilon_{\psi}))\ell^2}{96 (c+2)L^2}
 +O((\ell/L)^4) \Big].\nn
\eea
We see that both of them have nontrivial large $c$ corrections though with different structures.

\section{Conclusion and discussion}\label{sec6}

ETH is a fundamental issue in quantum thermodynamics and its validity for various situation should be scrutinized. An interesting version of ETH was proposed very recently in \cite{Lashkari:2016vgj,Dymarsky:2016aqv}, the so-called subsystem ETH which requires the difference between high energy state and microcanonical ensemble thermal state over all a local region should be exponentially suppressed by the entropy of the total system.  This can be further relaxed to the so-called subsystem weak ETH, which compares the high energy state and canonical ensemble thermal state. To be precise, the trace distance of the reduced density matrices should be power-law suppressed.

In this paper we check the validity of subsystem weak ETH for a 2D large $c$ CFT. We evaluate the R\'enyi entropy, entanglement entropy, and relative entropy of the reduced density matrices to measure the difference between the heavy primary state and thermal state.
We use these results and some information inequalities to get the bounds for the trace distance in large $c$ limit, and find
\be\label{trace d-1}
 \f{\mO(c^0)}{\log d} \leq t \leq \mO(c^0).
\ee
The upper bound is trivial. The lower bound depends on how the effective dimension $d$ of the subsystem scales with $c$, which is subtle to determine. Instead of using the relation \eq{trace d-1} as a definition of $d$, see, e.g., \cite{Dymarsky:2016aqv}, we treat it as an open issue and consider the possible interesting scenarios. One of these is that $d$  satisfies Fannes-Audenaert inequality and at the same time yields nontrivial lower bound of $t$. It is plausible that
\be \label{z59}
d \sim \ep^{\mO(c)}.
\ee
If this is true, the trace distance would be power-law suppressed so that it is consistent with the subsystem weak ETH.

We have to say we do not have a concrete proof that the result (\ref{z59}) is correct. The validity of the subsystem weak ETH really depends only on whether the large but finite effective dimension exists and if yes how it scales with the large central charge. It is an open question, and it is possible one has to calculate the trace distance explicitly to find the answer.

As pointed out in \cite{Lin:2016dxa}, the mismatch of the R\'enyi entropies of excited state and canonical ensemble thermal state at order $\ell^4$ originates from the mismatch of the one-point expectation values of the level 4 quasiprimary operator $\mA$. The same reason applies to the mismatch of entanglement entropies, and the non-vanishing of relative entropy at order $\ell^8$ in this paper. One possible resolution is that the excited state should be compared to the generalized Gibbs ensemble thermal state \cite{Rigol:2006,Mandal:2015jla,Cardy:2015xaa,Mandal:2015kxi,Vidmar:2016,deBoer:2016bov}, instead of the ordinary canonical thermal state. In fact, there are an infinite number of commuting conserved charges in the vacuum conformal family \cite{Sasaki:1987mm,Bazhanov:1994ft}, and in the generalized Gibbs ensemble one can also use noncommuting charges \cite{Cardy:2015xaa}. We will discuss about it in more details in a work that will come out soon \cite{work-disimilarity}.\footnote{We thank a JHEP referee for discussions about higher order conserved charges in KdV hierarchy and the generalized Gibbs ensemble.}

\section*{Acknowledgments}

We thank Huajia Wang for an initial participation of the project.
We would like to thank Pallab Basu, Diptarka Das, Shouvik Datta, Sridip Pal for their very helpful discussions.
We thank Matthew Headrick for his \emph{Mathematica} code \emph{Virasoro.nb}, which can be downloaded at \url{http://people.brandeis.edu/~headrick/Mathematica/index.html}.
JJZ would like to thank the organisers of \emph{The String Theory Universe Conference 2017} held in Milan, Italy on 20-24 February 2017 for being given the opportunity to present some of the results as gongshow and poster, and thank the participants, especially Geoffrey Comp\`ere, Debajyoti Sarkar and Marika Taylor, for stimulating discussions.
SH is supported by Max-Planck fellowship in Germany and the National Natural Science Foundation of China Grant No.~11305235.
FLL is supported by Taiwan Ministry of Science and Technology through Grant No.~103-2112-M-003-001-MY3 and No.~103-2811-M-003-024.
JJZ is supported by the ERC Starting Grant 637844-HBQFTNCER.

\appendix

\section{Some details of vacuum conformal family}\label{vcf}

We list the holomorphic quasiprimary operators in vacuum conformal family to level 8.
In level 2, we have the quasiprimary operator $T$, with the usual normalization $\a_T=\f{c}{2}$. In level 4, we have
\be \label{Adef}
\mA=(TT)-\f{3}{10}\p^2T, ~~ \a_{\mc A}=\f{c(5c+22)}{10}.
\ee
In level 6, we have the orthogonalized quasiprimary operators
\be
\mB=(\p T\p T)-\f{4}{5}(\p^2TT)-\f{1}{42}\p^4T,  ~~
\mD= \mC + \f{93}{70c+29} \mc B,
\ee
with the definition
\be
\mC = (T(TT))-\f{9}{10}(\p^2TT)-\f{1}{28}\p^4 T,
\ee
and the normalization factors are
\be
\a_{\mc B}=\frac{36c (70 c+29)}{175}, ~~
\a_{\mc D}=\frac{3 c (2 c-1) (5 c+22) (7 c+68)}{4 (70 c+29)}.
\ee
In level 8 we have the orthogonalized quasiprimary operators
\bea
&& \mc E= (\p^2 T\p^2 T)-\f{10}{9}(\p^3 T \p T)+\f{10}{63}(\p^4TT)-\f{1}{324}\p^6 T, \nn\\
&& \mc H=\mc F+ \frac{9 (140 c+83)}{50 (105 c+11)}\mc E,\\
&& \mc I=\mc G+ \frac{81 (35 c-51)}{100 (105 c+11)}\mc E + \frac{12(465 c-127)}{5 c(210 c+661)-251}\mc H, \nn
\eea
with the definitions
\bea
&& \mc F=(\p T(\p TT))-\f{4}{5}(\p^2 T(T T))+\f{2}{15}(\p^3 T \p T)-\f{3}{70}(\p^4TT),   \\
&& \mc G=(T(T(TT)))-\f{9}{5}(\p^2 T(TT))+\f{3}{10}(\p^3 T \p T)-\f{13}{70}(\p^4TT)-\f{1}{240}\p^6 T, \nn
\eea
and the normalization factors are
\bea
&& \a_{\mc E}=\frac{22880 c (105 c+11)}{1323}, \nn\\
&& \a_{\mc H}=\frac{26 c (5 c+22) (5 c (210 c+661)-251)}{125 (105 c+11)},\\
&& \a_{\mc I}=\frac{3 c (2 c-1) (3 c+46) (5 c+3) (5 c+22) (7 c+68)}{2 (5 c (210 c+661)-251)}.\nn
\eea


In this paper we need the structure constants
\be
C_{TTT}=c, ~~
C_{TT\mA}=\frac{c(5c+22)}{10},
\ee
and the four-point function on complex plane
\be
\lag T(z_1) T(z_{2}) T(z_{3}) T(z_{4}) \rag_\mC = \f{c^2}{4} \bigg( \f{1}{\lt( z_{12}z_{34}\rt)^4}
                                                       +\f{1}{\lt( z_{13}z_{24}\rt)^4}
                                                       +\f{1}{\lt( z_{14}z_{23} \rt)^4} \bigg)
+c \f{(z_{12}z_{34})^2+(z_{13}z_{24})^2+(z_{14}z_{23})^2}{(z_{12}z_{34}z_{13}z_{24} z_{14}z_{23})^2},
\ee
with the definitions $z_{ij} \equiv z_i-z_j$.


Under a general coordinate transformation $z\to f(z)$, we have the transformation rules
\bea \label{conftransf}
&& T(z)=f'^2 T(f)+\f{c}{12}s, ~~
   \mA(z)=f'^4\mA(f)+\f{5c+22}{30}s \Big( f'^2 T(f)+\f{c}{24}s \Big),                     \nn\\
&& \mB(z)= f'^6\mB(f)-\f{8}{5}sf'^4\mA(f)
          -\f{70c+29}{1050}sf'^4\p^2T(f)
          +\f{70c+29}{420}f'^2(f's'-2f''s)\p T(f)                                         \nn\\
&& \phantom{\mB(z)=}
          -\f{1}{1050}\lt( 28(5c+22)f'^2s^2+(70c+29)(f'^2s''-5f'f''s'+5f''^2s) \rt)T(f)   \nn\\
&& \phantom{\mB(z)=}
          -\f{c}{50400}\lt( 744s^3+ (70c+29)(4ss''-5s'^2) \rt),                           \nn\\
&& \mD(z)=f'^6\mD(f)+\f{(2c-1)(7c+68)}{70c+29}s \Big( \f{5}{4} f'^4\mA(f)+\f{5c+22}{48}s \big( f'^2T(f)+\f{c}{36}s \big) \Big),   \nn
\\
&& \mE(z) = f'^8\mE(f)
            -\frac{4510}{567}s f'^6 \mB(f)
            +\frac{50}{567} s f'^6 \p^2\mA(f)
            -\frac{25}{63} f'^4 (f' s'-2 s f'') \p\mA(f) \nn\\
&&\phantom{\mE(z) =}
            +\frac{4}{63}f'^2 (25 s f''^2+5 f'^2 s''+98 s^2 f'^2-25 f' f'' s') \mA(f)
            +\frac{105 c+11}{7938} s f'^6\p^4T(f) \nn\\
&&\phantom{\mE(z) =}
            -\frac{105 c+11}{1134}f'^4 (f' s'-2 s f'')\p^3T(f)
            +\frac{1}{5670}f'^2 \big(9 (105 c+11) (f'^2 s''+5 s f''^2-5 f' f'' s')  \nn\\
&&\phantom{\mE(z) =}
            +10 (120 c+77) s^2 f'^2\big)\p^2T(f)
            -\frac{1}{2268}\big((105 c+11) (2 s^{(3)} f'^3-30 s f''^3-18 f'^2 f'' s''+45 f' f''^2 s')  \nn\\
&&\phantom{\mE(z) =}
            + 10 (120 c+77) s f'^2 (f' s'-2 s f'')\big)\p T(f)
            +\frac{1}{79380 f'^2}\big(8 (3570 c+2629) s f'^4 s''\nn\\
&&\phantom{\mE(z) =} -5 (2940 c+2563) f'^4 s'^2+12 (1225 c+9449) s^3 f'^4  +700 (120 c+77) s f'^2 f'' (s f''-f' s')\nn\\
&&\phantom{\mE(z) =}
            +5 (105 c+11) (105 s f''^4+2 s^{(4)} f'^4-28 s^{(3)} f'^3 f''+126 f'^2 f''^2 s''-210 f' f''^3 s')\big)T(f) \nn\\
&&\phantom{\mE(z) =}
            +\frac{c}{952560}\big((105 c+11) (10 s s^{(4)}+63 s''^2-70 s^{(3)} s')
            +451 s (20 s s''-25 s'^2+52 s^3)\big), 
\eea\bea
&& \mH(z) = f'^8\mH(f)
           -\frac{8}{5} s f'^6\mD(f)
           +\frac{91 (5 c+22) (5 c (210 c+661)-251)}{540 (70 c+29) (105 c+11)}s f'^6\mB(f)  \nn\\
&& \phantom{\mH(z) =}
           -\frac{5 c (210 c+661)-251}{540 (105 c+11)}s f'^6\p^2\mA(f)
           +\frac{5 c (210 c+661)-251}{120 (105 c+11)}f'^4 (f' s'-2 s f'')\p\mA(f)  \nn\\
&& \phantom{\mH(z) =}
           +\frac{1}{150 (105 c+11)}f'^2 \big((5 c (210 c+661)-251) (5 f' f'' s'-5 s f''^2-f'^2 s'') \nn\\
&& \phantom{\mH(z) =}
           -8 (25 c (21 c+187)-951) s^2 f'^2\big)\mA(f)
           -\frac{(5 c+22) (5 c (210 c+661)-251)}{9000 (105 c+11)} s^2 f'^4\p^2T(f) \nn\\
&& \phantom{\mH(z) =}
           +\frac{(5 c+22) (5 c (210 c+661)-251)}{3600 (105 c+11)}s f'^2 (f' s'-2 s f'')\p T(f) \nn\\
&& \phantom{\mH(z) =}
           +\frac{5 c+22}{108000 (105 c+11)}\big(3 (5 c (210 c+661)-251) (-20 s^2 f''^2-8 s f'^2 s''+5 f'^2 s'^2+20 s f' f'' s') \nn\\
&& \phantom{\mH(z) =}
           -8 (15 c (210 c+2273)-7357) s^3 f'^2)T(f)
           -\frac{c (5 c+22) s}{1296000 (105 c+11)}(104 (465 c-127) s^3 \nn\\
&& \phantom{\mH(z) =}
           + 3 (5 c (210 c+661)-251) (4 s s''-5 s'^2)),\nn\\
&& \mI(z)=f'^8\mI(f) +
   \frac{(3 c+46) (5 c+3) (70 c+29)}{3 (5 c (210 c+661)-251)}s \Big(\mD(f) f'^6+\frac{5 (2 c-1) (7 c+68)}{8 (70 c+29)}s \Big( f'^4\mA(f)\nn\\
&& \phantom{\mI(z)=}
   +\frac{5c+22}{90} s \Big( f'^2 T(f) +\frac{c}{48} s\Big)\Big)\Big). \nn
\eea
In the above equations, we have the definition of Schwarzian derivative
\be
s(z)=\f{f'''(z)}{f'(z)}-\f32 \bigg( \f{f''(z)}{f'(z)} \bigg)^2,
\ee
and the shorthand notations
\be
f' \equiv f'(z), ~~
f'' \equiv f''(z), ~~
s \equiv s(z), ~~
s' \equiv s'(z), ~~
s'' \equiv s''(z), ~~
s^{(3)} \equiv s^{(3)}(z), ~~
\cdots.
\ee

For a general primary operator $\phi$ with conformal weight $h_\phi$ and normalization factor $\a_\phi=1$, we have the structure constants
\bea \label{Cphiphik}
&& C_{\phi\phi T}=h_\phi, ~~
   C_{\phi\phi\mA}=\f{h_\phi(5h_\phi+1)}{5}, ~~
   C_{\phi\phi\mB}=-\frac{2 h_{\phi}(14 h_{\phi}+1)}{35} ,  \nn\\
&& C_{\phi\phi\mD}=\frac{h_{\phi} [(70 c+29) h_{\phi}^2+(42 c-57) h_{\phi}+(8 c-2)]}{70 c+29}, ~~
   C_{\phi\phi\mE}=\frac{4 h_{\phi} (27 h_{\phi}+1)}{63},\nn\\
&& C_{\phi\phi\mH}=-\frac{2 h_{\phi} \big(10 (105 c+11) h_{\phi}^2+(435 c-218) h_{\phi}+55 c-4\big)}{25 (105 c+11)},\\
&& C_{\phi\phi\mI}=\frac{h_{\phi}}{5 c (210 c+661)-251}\big((5 c (210 c+661)-251) h_{\phi}^3+6 (c (210 c-83)+153) h_{\phi}^2 \nn\\
&& \phantom{C_{\phi\phi\mI}=} +(c (606 c-701)-829) h_{\phi}+6 c (18 c+13)-6\big).\nn
\eea
For a general holomorphic quasiprimary operator $\mX$, the non-vanishing of $C_{\phi\phi\mX}$ require that $\mX$ is bosonic and its conformal dimension $h_\mX$ is an even integer, and this leads to $C_{\phi\phi\mX}=C_{\phi\mX\phi}$. We have the three-point function on complex plane
\be
\lag \phi(\inf)\p^r\mX(z)\phi(0)\rag_\mC = \f{(-)^r(h_\mX+r-1)!}{(h_\mX-1)!} \f{C_{\phi\phi\mX}}{z^{h_\mX+r}}.
\ee
For a general operator $\mX$, we denote its expectation value on a cylinder with spatial period $L$ in excited state $|\phi\rag$ as $\lag\mX\rag_\phi=\lag\phi|\mX|\phi\rag_\cyl$. From translation symmetry in both directions of the cylinder, we know that $\lag\mX\rag_\phi$ is a constant. So for $r\in \Z$ and $r>0$, we have
\be \label{e17}
\lag\p^r\mX\rag_\phi=0.
\ee
By mapping the cylinder to a complex plane, using (\ref{conftransf}) and (\ref{Cphiphik}) we get the expectation values
\bea \label{phikphi}
&&\hspace{-7mm}
  \lag T \rag_\phi = \frac{\pi^2 (c-24 h_\phi)}{6 L^2}, ~~
  \lag\mA \rag_\phi = \frac{\pi^4  (c (5 c+22) -240 (c+2) h_\phi +2880 h_\phi^2 )}{180 L^4}, ~~
  \lag\mB \rag_\phi = -\frac{2 \pi^6  (31 c - 504 h_\phi )}{525 L^6}, \nn\\
&&\hspace{-7mm}
  \lag\mD \rag_\phi = \frac{\pi^6}{216 (70 c+29) L^6}\big(  c (2 c-1) (5 c+22) (7 c+68)
                                                    -72 (70 c^3+617c^2+938c-248) h_{\phi} \nn\\
&&\hspace{-7mm} \phantom{ \lag \mD \rag_\phi =}
                                                    +1728 (c+4) (70 c+29) h_{\phi}^2
                                                    -13824 (70 c+29) h_{\phi}^3 \big), \nn\\
&& \hspace{-7mm}
  \lag\mE\rag_\phi = \frac{572 \pi^8 (41 c-480 h_{\phi})}{59535 L^8}, \\
&& \hspace{-7mm}
  \lag\mH\rag_\phi = -\frac{13 \pi^8}{10125 (105 c+11) L^8}\big(c (5 c+22) (465 c-127)
                                                               -480 (195 c^2+479c-44) h_{\phi}
                                                               +8640 (105 c+11) h_{\phi}^2\big), ~~ \nn\\
&& \hspace{-7mm}
   \lag\mI\rag_\phi = \frac{\pi^8}{1296 (1050 c^2+3305 c-251) L^8} \big(
                        c (2 c-1) (3 c+46) (5 c+3) (5 c+22) (7 c+68) \nn\\
&& \hspace{-7mm}\phantom{\lag\mI\rag_\phi =}
                       -96 (1050 c^5+23465 c^4+153901 c^3+274132 c^2+22388 c-6864) h_{\phi} \nn\\
&& \hspace{-7mm}\phantom{\lag\mI\rag_\phi =}
                       +3456 (1050 c^4+16325 c^3+69963 c^2+65686 c-648) h_{\phi}^2  \nn\\
&& \hspace{-7mm}\phantom{\lag\mI\rag_\phi =}
                       -55296 (c+6) (1050 c^2+3305 c-251) h_{\phi}^3
                       +331776 (1050 c^2+3305 c-251) h_{\phi}^4 \big).\nn
\eea

\section{OPE of twist operators}\label{opests}

We review OPE of twist operators in the $n$-fold CFT that is denoted as $\CFT^n$ \cite{Headrick:2010zt,Calabrese:2010he,Chen:2013kpa,Chen:2013dxa}. We also define and calculate $C_{\Phi\Phi K}$, $\lag\Phi_K\rag_\Phi$, and $b_K$ that would be useful to subsection~\ref{sec3.1}. Note that in this paper we only consider contributions of the holomorphic sector. The twist operators $\s$ and $\td\s$ are primary operators with conformal weights\cite{Calabrese:2004eu}
\be
h_\s=h_{\td \s}=\f{c(n^2-1)}{24n}.
\ee
We have the OPE of twist operators \cite{Headrick:2010zt,Calabrese:2010he,Chen:2013kpa}
\be \label{sts}
\s(z)\td \s(w)=\f{c_n}{(z-w)^{2h_\s}} \sum_K d_K \sum_{r=0}^\inf \f{a_K^r}{r!}(z-w)^{h_K+r} \p^r\Phi_K(w).
\ee
Here $c_n$ is the normalization factor. The summation $K$ is over each orthogonalized holomorphic quasiprimary operator $\Phi_K$ in $\CFT^n$, and $h_K$ is the conformal weight of $\Phi_K$. We have definition
\be
a_K^r \equiv \f{C_{h_K+r-1}^r}{C_{2h_K+r-1}^r},
\ee
with $C_x^y$ denoting the binomial coefficient that is also written as $x \choose y$. To level 8, the $\CFT^n$ holomorphic quasiprimary operators has been constructed in \cite{Chen:2013dxa}, and we just list them in table~\ref{cftnqp}. The normalization factors $\a_K$ and OPE coefficients $d_K$ for all these quasiprimary operators can also be found in \cite{Chen:2013dxa}.

\begin{table}[htbp]
  \centering
\begin{tabular}{|c|c|c|c|c|c|c|c|c|c|c|}
  \cline{1-2}\cline{4-5}\cline{7-8}\cline{10-11}
  level            & operator && level            & operator     && level            & operator            && level & operator             \\\cline{1-2}\cline{4-5}\cline{7-8}\cline{10-11}
  0                & 1        && \multirow{4}*{6} & $\mB$, $\mD$ && \multirow{2}*{7} & $\mM$               &&       & $\mO$                \\\cline{1-2}\cline{5-5}\cline{8-8}\cline{11-11}
  2                & $T$      &&                  & $T\mA$       &&                  & $T\mJ$, $\mN$       &&       & $\mP$                \\\cline{1-2}\cline{5-5}\cline{7-8}\cline{11-11}
  \multirow{2}*{4} & $\mA$    &&                  & $\mK$        &&                  & $\mE$, $\mH$, $\mI$ && 8     & $TT\mA$              \\\cline{2-2}\cline{5-5}\cline{8-8}\cline{11-11}
                   & $TT$     &&                  & $TTT$        && 8                & $T\mB$, $T\mD$      &&       & $T\mK$, $\mQ$, $\mR$ \\\cline{1-2}\cline{4-5}\cline{8-8}\cline{11-11}
  5                & $\mJ$    && 7                & $\mL$        &&                  & $\mA\mA$            &&       & $TTTT$               \\
  \cline{1-2}\cline{4-5}\cline{7-8}\cline{10-11}
\end{tabular}
  \caption{To level 8, the $\CFT^n$ holomorphic quasiprimary operators. We have omitted the replica indices and their constraints. The definitions of $\mJ$, $\mK$, $\mL$, $\mM$, $\mN$, $\mO$, $\mP$, $\mQ$, $\mR$ can be found in \cite{Chen:2013dxa}, and the normalization factors $\a_K$ and OPE coefficients $d_K$ for all these quasiprimary operators can also be found therein.}\label{cftnqp}
\end{table}

From a holomorphic primary operator $\phi$ with normalization $\a_\phi=1$ in the original CFT, we can define the $\CFT^n$ primary operator
\be
\Phi = \prod_{j=0}^{n-1}\phi_j.
\ee
In subsection~\ref{sec3.1}, we need structure constant $C_{\Phi\Phi K}$ for the quasiprimary operators $\Phi_K$ in table~\ref{cftnqp}. The results can be written in terms of (\ref{Cphiphik}). First of all it is easy to see
\bea \label{CPhiPhiK1}
&& C_{\Phi\Phi T} = C_{\phi\phi T}, ~~
   C_{\Phi\Phi\mA} = C_{\phi\phi\mA}, ~~
   C_{\Phi\Phi\mB} = C_{\phi\phi\mB}, ~~
   C_{\Phi\Phi\mD} = C_{\phi\phi\mD}, \nn\\
&& C_{\Phi\Phi\mE} = C_{\phi\phi\mE}, ~~
   C_{\Phi\Phi\mH} = C_{\phi\phi\mH}, ~~
   C_{\Phi\Phi\mI} = C_{\phi\phi\mI}, ~~
   C_{\Phi\Phi TT} = C_{\phi\phi T}^2, \\
&& C_{\Phi\Phi T\mA} = C_{\phi\phi T}C_{\phi\phi\mA}, ~~
   C_{\Phi\Phi T\mB} = C_{\phi\phi T}C_{\phi\phi\mB}, ~~
   C_{\Phi\Phi T\mD} = C_{\phi\phi T}C_{\phi\phi\mD}, \nn\\
&& C_{\Phi\Phi \mA\mA} = C_{\phi\phi\mA}^2, ~~
   C_{\Phi\Phi TTT} = C_{\phi\phi T}^3, ~~
   C_{\Phi\Phi TT\mA} = C_{\phi\phi T}^2 C_{\phi\phi\mA}, ~~
   C_{\Phi\Phi TTTT} = C_{\phi\phi T}^4. \nn
\eea
There are vanishing structure constants
\be \label{CPhiPhiK2}
C_{\Phi\Phi\mJ}=C_{\Phi\Phi\mL}=C_{\Phi\Phi\mM}=C_{\Phi\Phi T\mJ}=C_{\Phi\Phi\mN}=0.
\ee
For $\mK$, $\mO$, and $\mP$ we have
\be \label{CPhiPhiK3}
C_{\Phi\Phi\mK}= -\frac{4}{5}C_{\phi\phi T}^2, ~~
C_{\Phi\Phi\mO}= -\frac{56}{45}C_{\phi\phi T}C_{\phi\phi\mA}, ~~
C_{\Phi\Phi\mP}= \frac{12}{7}C_{\phi\phi T}^2.
\ee
Finally, we have
\be \label{CPhiPhiK4}
C_{\Phi\Phi T\mK} = C_{\phi\phi T}C_{\phi\phi\mK}, ~~
C_{\Phi\Phi\mQ} = \f79 C_{\Phi\Phi T\mK}, ~~
C_{\Phi\Phi\mR} = \f{7}{11} C_{\Phi\Phi T\mK}.
\ee

It is easy to get $\lag\Phi_K\rag_\Phi$ that appear in (\ref{e16}) in  terms of (\ref{phikphi})
\bea\label{PhiKPhi1}
&& \lag T \rag_\Phi = \lag T \rag_\phi, ~~
   \lag \mA \rag_\Phi = \lag \mA \rag_\phi, ~~
   \lag \mB \rag_\Phi = \lag \mB \rag_\phi, ~~
   \lag \mD \rag_\Phi = \lag \mD \rag_\phi, ~~
   \lag \mE \rag_\Phi = \lag \mE \rag_\phi, ~~
   \lag \mH \rag_\Phi = \lag \mH \rag_\phi, \nn\\
&& \lag \mI \rag_\Phi = \lag \mI \rag_\phi, ~~
   \lag TT \rag_\Phi = \lag T \rag_\phi^2, ~~
   \lag T\mA \rag_\Phi = \lag T \rag_\phi \lag \mA \rag_\phi, ~~
   \lag T\mB \rag_\Phi = \lag T \rag_\phi \lag \mB \rag_\phi, ~~
   \lag T\mD \rag_\Phi = \lag T \rag_\phi \lag \mD \rag_\phi, \nn\\
&& \lag \mA\mA \rag_\Phi = \lag \mA \rag_\phi^2, ~~
   \lag TTT \rag_\Phi = \lag T \rag_\phi^3, ~~
   \lag TT\mA \rag_\Phi = \lag T \rag_\phi^2 \lag \mA \rag_\phi, ~~
   \lag TTTT \rag_\Phi = \lag T \rag_\phi^4.
\eea
Because of (\ref{e17}) we have the vanishing results
\be\label{PhiKPhi2}
\lag \mJ \rag_\Phi =
\lag \mK \rag_\Phi =
\lag \mL \rag_\Phi =
\lag \mM \rag_\Phi =
\lag \mN \rag_\Phi =
\lag \mO \rag_\Phi =
\lag \mP \rag_\Phi =
\lag \mQ \rag_\Phi =
\lag \mR \rag_\Phi =
\lag T\mJ \rag_\Phi =
\lag T\mK \rag_\Phi =0.
\ee

From OPE coefficient $d_K$ for quasiprimary operators in table~\ref{cftnqp} we may define $b_K$ by summing over the indices of $\Phi_K$
\be
b_K = \sum_{j_1,\cdots} d_K^{j_1\cdots}.
\ee
For examples, in table~\ref{cftnqp} $T$ denotes operators $T_j$ with $0\leq j \leq n-1$, and $TT\mA$ denotes operators $T_{j_1}T_{j_2}\mA_{j_3}$ with $0\leq j_{1,2,3} \leq n-1$ and the constraints $j_1 < j_2$, $j_1\neq j_3$, $j_2 \neq j_3$, and so we have
\be
b_T = \sum_{j=0}^{n-1}d_T=n d_T,
\ee
and
\be
b_{TT\mA} = \sum_{0\leq j_{1,2,3} \leq n-1}d_{TT\mA}^{j_1j_2j_3}~\textrm{with constraints}~j_1 < j_2,~j_1\neq j_3,~j_2 \neq j_3.
\ee
Using the results of $d_K$ and the summation formulas in \cite{Chen:2013dxa} we get the $b_K$ we need. In subsection~\ref{sec3.1} we need
\bea \label{bK1}
&& b_T=\frac{n^2-1}{12 n}, ~~
   b_\mA=\frac{(n^2-1)^2}{288 n^3}, ~~
   b_\mB=-\frac{(n^2-1)^2 (70 c n^2+122 n^2-93)}{10368 (70 c+29) n^5}, \nn\\
&& b_\mD=\frac{(n^2-1)^3}{10368 n^5}, ~~
   b_\mE=\f{(n^2-1)^2 (11340 c n^4+11561 n^4-16236 n^2+5863)}{65894400 (105 c+11) n^7}, \nn\\
&& b_\mH=-\f{(n^2-1)^3 \big(3150 c^2 n^2+c (15960 n^2-6045)-2404 n^2+1651\big)}{539136 (5 c (210 c+661)-251) n^7}, ~~
   b_\mI=\frac{(n^2-1)^4}{497664 n^7}, \nn\\
&& b_{TT}=\frac{(n^2-1) (5 c (n+1) (n-1)^2+2 n^2+22)}{1440 c n^3} , ~~
   b_{T\mA}=\frac{(n^2-1)^2 (5 c (n+1) (n-1)^2+4 n^2+44)}{17280 c n^5}, \nn\\
&& b_{T\mB}=-\frac{(n^2-1)^2}{13063680 c (70 c+29) n^7}\big( 7350 n^2 c^2 (n-1)^2 (n+1)\nn\\
&& \phantom{b_{T\mB}=}
                                                            +35 c (366 n^5-238 n^4-645 n^3+2369 n^2+279 n-403)
                                                            +2 (6787 n^4+71089 n^2-65348) \big), \nn\\
&& b_{T\mD}=\frac{(n^2-1)^3 \big(5 c (n+1) (n-1)^2+6 n^2+66\big)}{622080 c n^7}, \\
&& b_{\mA\mA}=\frac{1}{5806080 c (5 c+22) n^7}( 175 c^2 (n+1)^4 (n-1)^5\nn\\
&& \phantom{b_{\mA\mA}=}
                                               +70 c (n^2-1)^3 (11 n^3-7 n^2-11 n+55)
                                               +8 (n^2-1)(n^2+11) (157 n^4-298 n^2+381)), \nn\\
&& b_{TTT}=\frac{(n-2) (n^2-1)}{362880 c^2 n^5} \big(35 c^2 (n+1)^2 (n-1)^3+42 c (n^4+10 n^2-11)-16 (n+2) (n^2+47)\big), \nn\\
&& b_{TT\mA}=\f{(n-2) (n^2-1)}{14515200 c^2 n^7}\big( 175 c^2 (n+1)^3 (n-1)^4
                                                     +350 c (n^2-1)^2 (n^2+11) \nn\\
&& \phantom{b_{TT\mA}=}
                                                     -128 (n+2) (n^4+50 n^2-111) \big), \nn\\
&& b_{TTTT}=\frac{(n-3) (n-2) (n^2-1)}{87091200 c^3 n^7}\big( 175 c^3 (n+1)^3 (n-1)^4
                                                             +420 c^2 (n^2-1)^2 (n^2+11) \nn\\
&& \phantom{b_{TTTT}=}
                                                             -4 c (59 n^5+121 n^4+3170 n^3+6550 n^2-6829 n-11711)
                                                             +192 (n+2) (n+3) (n^2+119) \big). \nn
\eea
In subsection~\ref{sec3.1}, we also need
\bea \label{bK2}
&& b_\mK=-\frac{(n^2-1)\big(70 c (n-1)^2 (n+1) n^2-2 n^4+215 n^2-93\big)}{725760 c n^5}, \nn\\
&& b_\mO=-\f{(n^2-1)^2\big(210 c (n-1)^2 (n+1) n^2+38 n^4+1445 n^2-403\big)}{37739520 c n^7},\nn\\
&& b_\mP=\f{(n^2-1)\big(11340 c (n-1)^2 (n+1) n^4-1481 n^6+27797 n^4-22099 n^2+5863\big)}{6918912000 c n^7},\\
&& b_{T\mK}+\f79 b_{\mQ}+\f7{11}b_{\mR}=-\frac{(n-2) (n^2-1)}{188697600 c^2 n^7}
                                      \big( 1050 c^2 (n+1)^2 n^2 (n-1)^3   \nn\\
&& \phantom{b_{T\mK}+\f79 b_{\mQ}+\f7{11}b_{\mR}=}
                                           +5 c (n^2-1)(122 n^4+2369 n^2-403)
                                           -4 (n+2) (81 n^4+4600 n^2-2041) \big). \nn
\eea

\section{Useful summation formulas}\label{usf}

Most of the summation formulas that are used in this paper can be found in \cite{Chen:2013dxa}. There are two other ones
\bea
&& \sum_{\neq} \f{1}{s_{j_1j_2}^2 s_{j_2j_3}^4 s_{j_3j_1}^4} = \frac{4 n (n^2-4) (n^2-1) (n^2+19) (n^4+19 n^2+628)}{467775},  \nn\\
&&  \sum_{\neq} \f{c_{j_1j_2} c_{j_1j_3}}{s_{j_1j_2}^3 s_{j_1j_3}^3 s_{j_2j_3}^4} = \frac{2 n (n^2-25) (n^2-4) (n^2-1) (n^4+30 n^2+419)}{467775},
\eea
and the summations are in the range $0\leq j_{1,2,3} \leq n-1$ with the constraints $j_1\neq j_2$, $j_1\neq j_3$ and $j_3\neq j_1$. Here we have used the shorthand $s_{j_1j_2}=\sin\f{\pi(j_1-j_2)}{n}$, $c_{j_1j_2}=\cos\f{\pi(j_1-j_2)}{n}$, and et al.

We define the summation of $k$ indices $0\leq j_{1,2,\cdots,k} \leq n-1$
\be
\sum_{\neq} f(j_1,j_2,\cdots,j_k),
\ee
with the constraints that any two of the indices are not equal and the function $f(j_1,j_2,\cdots,j_k)$ is totally symmetric for the $k$ arguments. First we have
\be
\sum_{\neq'} f(0,j_2,\cdots,j_k) = \f{1}{n} \sum_{\neq} f(j_1,j_2,\cdots,j_k),
\ee
with the summation $\neq'$ of the left-hand side being over $1\leq j_{2,\cdots,k} \leq n-1$ and the constraints that any two of the indices are not equal.
Then we have
\be
\sum_{\neq'} f(j_1,j_2,\cdots,j_k) = \f{n-k}{n} \sum_{\neq} f(j_1,j_2,\cdots,j_k),
\ee
with the summation of the left-hand side being over $1\leq j_{1,2,\cdots,k} \leq n-1$.

\providecommand{\href}[2]{#2}\begingroup\raggedright\endgroup


\end{document}